\documentclass[a4paper,11pt]{article}
\pdfoutput=1 

\usepackage{jheppub} 
\usepackage{graphicx,color}
\usepackage{enumitem}
\usepackage{braket}
\usepackage{slashed}
\usepackage[utf8]{inputenc}
\usepackage{hyperref}
\usepackage{float}
\usepackage{caption}
\usepackage{subfigure}
\usepackage{makecell}
\usepackage{epsfig}
\usepackage{amsmath,amssymb}
\usepackage{hyperref}
\usepackage{diagbox}
\usepackage{lscape}
\usepackage{wrapfig}
\usepackage{algorithm}
\usepackage[noend]{algpseudocode}
\usepackage{amsthm}
\usepackage{mathrsfs}

\hypersetup{
	colorlinks=true,
	citecolor=black,
	filecolor=black,
	linkcolor=blue,
	urlcolor=blue
}

\usepackage{dsfont}

\newcommand{\comments}[1]{}

\def\bel#1{\begin{equation} \label{#1}}

\def\ED3{{\scriptscriptstyle ED3}}

\newcommand{\beq}{\begin{equation}}  \newcommand{\eeq}{\end{equation}}
\newcommand{\bal}{\begin{aligned}}   \newcommand{\eal}{\end{aligned}}
\def\ov{\overline}
\newcommand{\bmat}{\left(\begin{array}}
\newcommand{\emat}{\end{array}\right)}

\newcommand{\cO}{\mathcal{O}}

\newcommand{\cK}{\mathcal{K}}
\newcommand{\cN}{\mathcal{N}}

\newcommand{\cI}{\mathcal{I}}

\newcommand{\cR}{\mathcal{R}}

\newcommand{\cM}{\mathcal M}

\newcommand{\ee}{\text{e}}

\usepackage{comment}
\usepackage[normalem]{ulem}

\def\bZ{\mathbb{Z}}
\newcommand{\I}{\text{i}}

\newcommand{\kom}{\, ,\quad }

\newcommand*{\dif}{{\,\rm d}}

\newcommand*{\p}{\mathop{}\!\mathrm \partial}

\newcommand{\bC}{\mathbb{C}}

\newcommand{\bR}{\mathbb{R}}

\title{Deep observations of the Type IIB flux landscape}

\author[a]{A. Chauhan,}
\author[b,c]{M. Cicoli,}
\author[d]{S. Krippendorf,}
\author[a]{A. Maharana,}
\author[b,c]{P. Piantadosi,}
\author[e,f]{\\A. Schachner}

\affiliation[a]{\footnotesize Harish-Chandra Research Institute, 
A CI of Homi Bhabha National Institute, Chhatnag Road, Jhunsi, Allahabad, India 211019.}
\affiliation[b]{\footnotesize Dipartimento di Fisica e Astronomia, Università di Bologna, via Irnerio 46, 40126 Bologna, Italy}
\affiliation[c]{\footnotesize INFN, Sezione di Bologna, viale Berti Pichat 6/2, 40127 Bologna, Italy}
\affiliation[d]{\footnotesize University of Cambridge, Cavendish Laboratory and DAMTP, Cambridge CB3 0WA, United Kingdom}
\affiliation[e]{\footnotesize ASC for Theoretical Physics, Ludwig-Maximilian-University, 80333 Munich, Germany}
\affiliation[f]{\footnotesize Department of Physics, Cornell University, Ithaca, NY 14853, USA}

\emailAdd{amanchauhan@hri.res.in}
\emailAdd{michele.cicoli@unibo.it}
\emailAdd{slk38@cam.ac.uk}
\emailAdd{anshumanmaharana@hri.res.in}
\emailAdd{pellegrin.piantados2@unibo.it}
\emailAdd{a.schachner@lmu.de}

\abstract{
We present deep observations in targeted regions of the string landscape through a combination of analytic and dedicated numerical methods. Specifically, we devise an algorithm designed for the systematic construction of Type IIB flux vacua in finite regions of moduli space. Our algorithm is universally applicable across Calabi-Yau orientifold compactifications and can be used to enumerate flux vacua in a region given sufficient computational efforts. As a concrete example, we apply our methods to a two-modulus Calabi-Yau threefold, demonstrating that systematic enumeration is feasible and revealing intricate structures in vacuum distributions. Our results highlight local deviations from statistical expectations, providing insights into vacuum densities, superpotential distributions, and moduli mass hierarchies. This approach opens pathways for precise, data-driven mappings of the string landscape, complementing analytic studies and advancing the understanding of the distribution of flux vacua. This allows us to obtain different types of solutions with hierarchical suppressions, e.g.~vacua with small values of the Gukov-Vafa-Witten superpotential $|W_0|$. We find an example with $|W_0| = 5.547 \times 10^{-5}$ at large complex structure, without light directions and the use of non-perturbative effects. 
}

\begin{document} 
\maketitle
\flushbottom

\section{Introduction}

The quest to quantitatively understand the low-energy limits of string theory holds significant promise for advancing Beyond the Standard Model (BSM) physics and precision holography. By providing insights into fine-tuning mechanisms and uncovering the structures underlying fundamental interactions, a robust grasp of the string landscape could guide theoretical predictions and phenomenological applications. However, the complexity of the string landscape raises a critical question: \emph{Can we actually zoom in on the string landscape and systematically explore viable regions in practice?}

To date, progress in exploring the string landscape has largely relied on searches in hand-selected examples or on statistical arguments employing suitable approximations. These approaches, while valuable, remain inherently incomplete for achieving a comprehensive phenomenological understanding. Hand-selected searches often overlook the full breadth of vacua, while statistical methods rest on untested assumptions, accessing only selected properties of vacua. Conjecture-driven approaches, though insightful, typically offer qualitative rather than quantitative assessments of the landscape's global structure. This paper addresses the need for a data-driven, exhaustive exploration of accessible regions within the landscape, moving toward a more complete and systematic understanding.

Focusing on Type IIB flux vacua \cite{Giddings:2001yu}, we aim to develop a precise and targeted deep observation of the string landscape. This setting, chosen for its computational control, enables us to meet specific phenomenological requirements, including constraints on $g_s$, $W_0$, and moduli masses. Dedicated numerical tools, particularly the \texttt{JAXVacua} framework developed in \cite{Dubey:2023dvu}, are leveraged to explore regions of moduli space exhaustively, providing new insights into the properties and distributions of vacua. These observations extend beyond phenomenological requirements, offering a versatile framework for probing generic properties excluded from previous models and inspiring model building through localised, deep studies.

Specifically, our investigations test predictions regarding the finiteness \cite{Acharya:2006zw,Grimm:2021vpn,Grimm:2023lrf} and density of string flux vacua, corroborating earlier works such as \cite{Douglas:2003um,Ashok:2003gk,Douglas:2004zg,Denef:2004ze}, which rest on the continuous flux approximation. For a given compactification manifold, the vastness of the landscape \cite{Bousso:2000xa} arises from multiple discrete flux configurations, with the number of vacua expected to scale as $Q_{D3}^{2h^{1,2}+2}$ \cite{Douglas:2003um,Ashok:2003gk,Douglas:2004zg,Denef:2004ze}, where $h^{1,2}$ is the number of complex structure moduli and $Q_{D3}$ denotes the maximum D3-charge set by the orientifold. Exhaustively searching for vacua at large $Q_{D3}$ or $h^{1,2}$ is practically infeasible. Instead, we focus on developing numerical methods to systematically generate solutions in specific regions of moduli space with desired properties. We accomplish this by deriving new bounds on the flux landscape that are crucial for systematising the generation of flux vacua, building upon and extending the findings of \cite{Plauschinn:2023hjw}. 

As an application of our methods, we study a two-moduli model: the degree 18 hypersurface in $\mathbb{CP}^4_{[1,1,1,6,9]}$ at its symmetric locus \cite{Giryavets:2003vd} at large complex structure \cite{Denef:2004dm}. Our findings reveal patterns in the distributions of $W_0$, $g_s$, moduli masses, and clustering within flux space, echoing structures observed in prior work \cite{Martinez-Pedrera:2012teo}. Comparing our results with statistical expectations \cite{Douglas:2003qc,Ashok:2003gk,Douglas:2004zg,Denef:2004ze} highlights the importance of algorithmic choices in landscape mapping. For example, discrepancies between our findings and \cite{Martinez-Pedrera:2012teo}, which differ by orders of magnitude, underscore the necessity of precise numerical approaches to capture nuanced structures in the flux landscape.

Through our investigation, we identified discrepancies between observed and predicted numbers of vacua in specific moduli space regions. While we observe global agreement with the predictions of \cite{Denef:2004ze}, local over- or under-estimates reveal discrepancies in the expected behaviour of the vacuum density. We identified distinctive patterns in the distribution of the superpotential $W_0$ in the complex plane. These distributions exhibit symmetries along certain axes but lack of angular symmetry. The appearance of circular structures and voids in the distribution of the axio-dilaton is partially explained by specific flux configurations. Additionally, the moduli mass distribution reveals a significant range of values with a clear hierarchy between minimal and maximal masses. Axionic and moduli directions also exhibit notable mixing. 

In a similar spirit, exhaustive explorations have been performed in other corners of the string landscape (e.g.~\cite{Lerche:1986cx,Cvetic:2001nr,Lebedev:2006kn,Faraggi:2006bc,Anderson:2013xka,Braun:2014lwp,Taylor:2015xtz,Cvetic:2019gnh,Becker:2022hse}) which have resulted in distinct vacua distributions. We stress that these approaches, in contrast to our one, did not require direct numerical optimisation, i.e.~to explicitly find vacuum solutions it was not necessary to solve an optimisation problem for multiple continuous moduli fields.

This paper is organised as follows. Sec.~\ref{sec:typeIIB} outlines conventions and reviews Type IIB flux compactifications. In Sec.~\ref{sec:bounds}, we discuss bounds on fluxes and the axio-dilaton critical value for the vacua search algorithm. Sec.~\ref{sec:algorithm} describes a general algorithm for obtaining flux vacua in targeted regions of moduli space. Sec.~\ref{sec:results} summarises our results and analysis of flux vacua, followed by conclusions in Sec.~\ref{sec:conclusions}. App.~\ref{sec:appendix} provides technical details on moduli space integrals used in the analysis. Our data will be made available on the following GitHub repository \href{https://github.com/ml4physics/JAXvacua}{\texttt{https://github.com/ml4physics/JAXvacua}}.

\section{Type IIB flux compactifications}
\label{sec:typeIIB}

In this section we briefly introduce Type IIB flux compactifications and gather results important for this work. This sets the conventions and notations that will be used throughout. For detailed reviews on the subject the reader can refer to \cite{Grana:2005jc, Douglas:2006es}. 

\subsection{Calabi-Yau compactifications at large complex structure}
\label{sec:lcs}

Let $(X_3,\widetilde{X}_3)$ be a pair of mirror dual Calabi-Yau threefolds and ${\mathcal{I}}:\, X_3\rightarrow X_3$ be a holomorphic and isometric involution of $X_3$ under which the holomorphic $3$-form transforms as $\Omega\mapsto -\Omega$. We then denote by $X_3/\mathcal{I}$ the corresponding O3/O7 orientifold on which we compactify Type IIB superstring theory. The resulting effective supergravity theory preserves $\mathcal{N}=1$ supersymmetry in four dimensions.

Under the orientifold action, the cohomology groups $H^{p,q}(X_3,{\mathcal{I}})$ split into odd and even eigenspaces, $H^{p,q}_{\pm}(X_3,\mathbb{Q})$. The complex structure moduli surviving this projection come in $\mathcal{N}=1$ chiral multiplets counted by $h^{1,2}_-(X_3,{\mathcal{I}})= \text{dim}\, (H^{1,2}_-(X_3,\mathbb{Q}))$ and will be denoted by $z^i$, $i=1,\ldots,h^{1,2}_-(X_3,{\mathcal{I}})$. In this work, we remain agnostic about other moduli sectors, see however \cite{Cicoli:2013cha,Demirtas:2021nlu,McAllister:2024lnt} for attempts to stabilise all moduli in similar setups. For simplicity, we assume that $h^{1,2}_+(X_3,{\mathcal{I}})=0$ such that $h^{1,2}_-(X_3,{\mathcal{I}})=h^{1,2}(X_3)$.\footnote{Orientifolds with these properties can e.g.~be obtained systematically using techniques described in \cite{Moritz:2023jdb}, see also \cite{Jefferson:2022ssj}.}

Next, we introduce a symplectic basis of $\{\Sigma_{I},\Sigma^I\} \subset H_3(X_3,\mathbb{Z})$ together with the corresponding Poincaré dual forms $\{\alpha^I,\beta_I\}$. We then define the \emph{periods} by integrating the holomorphic $3$-form $\Omega$ over these cycles, and collect them in the period vector $\Pi$, that is,
\begin{equation}
\label{eq:PeriodVecGen}
X^I=\int_{\Sigma_{I}}\Omega=\int_{X_3} \Omega\wedge \alpha^I\, ,\quad \mathcal{F}_I=\int_{\Sigma^I}\Omega=\int_{X_3} \Omega \wedge \beta_I \, , \quad \Pi =\left (\begin{array}{c} \mathcal{F}_I \\ X^I \end{array}\right ) \, .
\end{equation}
The periods $X^I$ serve as homogeneous complex coordinates on a local patch of the complex structure moduli space of $X_3$. Away from the locus $X^0=0$, we introduce projective coordinates $z^i =X^i/X^0$, $i=1,\ldots,h^{1,2}(X_3)$, and normalise $\Omega$ such that $X^0=1$. The dual periods $\mathcal{F}_I=\mathcal{F}_I(z)$ are then determined by a prepotential $F(z)$ through
\begin{equation}
    \mathcal{F}_i(z)=\partial_{z^i} F(z) \, ,\quad \mathcal{F}_0 =2F-z^i\p_{z^i}F\, .
\end{equation}
To compute the periods entering the GVW superpotential \eqref{eq:SupPotPerVec}, we focus on \emph{Large Complex Structure} (LCS) regions of the complex structure moduli space $\cM_{\text{cs}}(X_3)$. Mirror symmetry maps the LCS region of Type IIB reduced on $X_3$ to the \emph{large volume region} of the Type IIA compactified on the mirror dual CY $\widetilde{X}_{3}$. Thus, using mirror symmetry, one can show that the prepotential $F(z)$ at LCS takes the form \cite{Candelas:1990rm,Ceresole:1992su, Candelas:1993dm,Hosono:1993qy,Hosono:1994av}
\begin{equation}
    \label{eq:prepotential}
    F(z)=-\frac{1}{6}\widetilde{\kappa}_{ijk}\, z^i\,z^j\,z^k+\frac{1}{2}a_{ij}\,z^i\,z^j+b_i\,z^i +\tilde{\xi} + F_{\text{inst}}(z).
\end{equation}
Here, $\widetilde{\kappa}_{ijk}$ are the triple intersection numbers of $\widetilde{X}_{3}$. Various parameters appearing in \eqref{eq:prepotential} are given in terms of the $(1,1)$-forms $J_{i}\in H^{1,1}(\widetilde{X}_{3},\bZ)$ and the second Chern class of the mirror manifold $\widetilde{X}_{3}$ denoted by $c_2(\widetilde{X}_{3})$, as follows
\begin{align}
\widetilde{\kappa}_{ijk} &= \int_{\widetilde{X}_{3}} \, J_i \wedge J_j \wedge J_k\kom a_{ij} = \frac{1}{2}\int_{\widetilde{X}_{3}} \, J_i \wedge J_j \wedge J_j\,\text{mod}\,\mathbb{Z}\; ,\quad \nonumber\\
b_j &= \frac{1}{4!}\int_{\widetilde{X}_{3}} \,c_2(\widetilde{X}_{3}) \wedge J_j\kom \tilde{\xi} = \frac{\I}{2}\frac{\zeta(3)\, \chi(\widetilde{X}_{3})}{(2\pi)^3}\,.
\end{align}
The non-perturbative contributions $F_{\text{inst}}$ in \eqref{eq:prepotential} arise from worldsheet instanton effects on the mirror dual side, and are given by~\cite{Hosono:1994av, Hosono:1994ax}
\begin{equation}
\label{eq:InstCorrections} 
    F_{\text{inst}}(z)=-\dfrac{1}{(2\pi \I)^{3}}\sum_{q \in \cM(\widetilde{X}_{3})}\, \mathscr{N}_{\tilde{\mathbf{q}}}\, \text{Li}_{3}\left (\ee^{2\pi \I\, q_i\, z^i}\right )\kom \text{Li}_{3}(x)=\sum_{m=1}^{\infty}\, \dfrac{x^{m}}{m^{3}}\, .
\end{equation}
Here, the sum runs over the effective curves $q$ in the \emph{Mori cone} $\cM(\widetilde{X}_{3})$ of the mirror $\widetilde{X}_{3}$ and $\mathscr{N}_{\tilde{\mathbf{q}}}$~\cite{Gopakumar:1998ii,Gopakumar:1998jq} are the genus-zero \emph{Gopakumar-Vafa (GV) invariants}.  
A systematic procedure for evaluating these invariants was developed by HKTY~\cite{Hosono:1993qy, Hosono:1994ax}. In practice, they can be computed using the software package \texttt{CYTools}~\cite{Demirtas:2022hqf, Demirtas:2023als}. The validity of the ansatz \eqref{eq:prepotential} for $F$ is restricted to the region where the LCS expansion converges~\cite{Hosono:1994av}, see also \cite{Candelas:1994hw, Klemm:1999gm}. In particular, the imaginary parts of the complex structure moduli $z^i$ take values inside the K\"ahler cone $\cK_{\widetilde{X}_{3}}$ of $\widetilde{X}_{3}$ defined as
\begin{equation}
\label{eq:DefKahlerCone}
    \cK_{\widetilde{X}_{3}} = \lbrace J\in H^{1,1}(\widetilde{X}_{3},\bR):\, \text{Vol}_{J}(U)>0\; \forall \text{ sub-varieties }U\rbrace\, .
\end{equation}
Here, the sub-varieties consist of effective curves, effective divisors, and $\widetilde{X}_{3}$ itself. This describes the moduli space of Kähler structures on $\widetilde{X}_{3}$, parametrised by a K\"ahler form $J$. In practice the K\"ahler cone computations are performed using \texttt{CYTools}. 

\subsection{Flux superpotential and vacua}
\label{sec:flux_vacua}

Let us now turn on background fluxes for the ten-dimensional gauge fields $H_3$ and $F_3$ along the compact directions. In terms of the above symplectic basis, we introduce the flux quanta
\begin{equation}
(f_2)^I 
=\int_{\Sigma_{I}}F_3\, ,\quad (f_1)_I
=\int_{\Sigma^I}F_3\, ,\quad (h_2)^I
=\int_{\Sigma_{I}}H_3\, ,\quad (h_1)_I
=\int_{\Sigma^I}H_3\, ,
\end{equation}
and collect them in two integral flux vectors $f,h\in \mathbb{Z}^{2(h^{2,1}+1)}$
\begin{equation}\label{eq:fluxdef}
    f=\left (\begin{array}{c}
                f_{1} \\ 
                f_{2}
                \end{array} 
    \right ) \, ,\quad 
    h=\left (\begin{array}{c}
                h_{1}\\ 
                h_{2} 
            \end{array} 
    \right )  \kom f_{1},f_{2},h_{1},h_{2}\in\mathbb{Z}^{h^{2,1}+1}\, .
\end{equation}
These fluxes are constrained by Gauss's law for the ten-dimensional gauge fields, which reads
\begin{equation}
     2\left(N_{\text{D3}}-N_{\overline{\text{D3}}}\right) + N_{\text{flux}} - Q_{D3} = 0\,,\label{eq:D3tadpole}
\end{equation} 
where $N_{\text{D3}}$ ($N_{\overline{\text{D3}}}$) is the number of spacetime-filling (anti-)D3-branes. Further, we introduced
\begin{equation}\label{eq:fluxtadpole}
     Q_{D3} =  \dfrac{\chi_f}{2}\, ,\quad N_{\text{flux}} =  \int_X H_3 \wedge F_3 = f^{\, T}\cdot \Sigma\cdot h\,,
\end{equation} 
in terms of the Euler character $\chi_f$ of the fixed locus of $\mathcal{I}$ in $X_3$. The D3-tadpole cancellation condition \eqref{eq:D3tadpole} has to be satisfied in any consistent solution of string theory. Thus, if e.g. the D3-charge contribution from fluxes is such that $N_{\text{flux}}<Q_{D3}$, one needs to add spacetime-filling D3-branes. 

In the four-dimensional $\cN=1$ supergravity theory, the tree-level K\"ahler potential $K$ for the complex structure moduli and the axio-dilaton is
\begin{equation}
\label{eq:TreeLevKP}
    K=-\ln( -\I\, \Pi^{\dagger}\cdot\Sigma\cdot \Pi ) - \ln\left(-\I({\tau}-\ov{\tau})\right) \kom \Sigma=\left (\begin{array}{cc}
        0 & \mathds{1} \\ [-0.2em]
        -\mathds{1} & 0
        \end{array} \right )\, .
\end{equation}
The $F$-term scalar potential is given by $V_F= V_{\rm flux}/\mathcal{V}^2$ where $\mathcal{V}$ is the dimensionless Calabi-Yau volume in string units, while the flux potential reads
\begin{equation}
\label{eq:scalarpotential}
    V_{\text{flux}}= \ee^{K} \left ( K^{\tau\bar{\tau}}D_{\tau} W\, D_{\bar{\tau}}\overline{ W}+ K^{i\bar\jmath}D_{i} W\, D_{\bar\jmath}\overline{ W}\right )\, ,\quad D_I W = \partial_I W + (\partial_{I}K) W
\end{equation}
where $W$ is the Gukov-Vafa-Witten (GVW) superpotential~\cite{Gukov:1999ya},
\begin{equation}
\label{eq:SupPotPerVec} 
    W=\int_{X_3} G_{3} \wedge \Omega = \left (f-\tau h\right )^{T}\cdot \Sigma\cdot   \Pi(z) \, .
\end{equation}
While $W$ is protected by non-renormalisation theorems against perturbative corrections \cite{Burgess:2005jx}, it receives non-perturbative contributions from D-brane instantons which we ignore subsequently. Moreover, we also ignore perturbative corrections to the K\"ahler potential since they are expected to be subdominant in the LCS regime and when the string coupling is small.

The action of Type IIB superstring theory enjoys an $\text{SL}(2,\mathbb{Z})$ symmetry under which the axio-dilaton and $3$-form fluxes transform as
\begin{equation}
\tau \to  {{a \tau + b} \over {c\tau+ d}} \, ,\quad \begin{pmatrix}
h \cr f
\end{pmatrix}
\to
\begin{pmatrix}
d & c \cr
b & a
\end{pmatrix}
\begin{pmatrix}
 h \cr f
\end{pmatrix} \, ,\quad \begin{pmatrix}
a & b \cr
c & d
\end{pmatrix} \in SL(2,\mathbb{Z})\,.   
\label{eq:SLtrafo}   
\end{equation}
Under this transformation, the tadpole \eqref{eq:fluxtadpole} remains invariant, but the GVW superpotential \eqref{eq:SupPotPerVec} transforms non-trivially.
By performing $\text{SL}(2,\mathbb{Z})$ transformations successively, $\tau=c_0 + \I s$ takes values in a \emph{fundamental domain} $\cM_\tau$ of $\text{SL}(2,\mathbb{Z})$ which we choose as
\begin{equation}\label{eq:FDAD}
   \cM_\tau = \biggl \{\tau=c_0+\I s\in\bC:\, |c_{0}|\leq 0.5\kom \dfrac{\sqrt{3}}{2}\leq s\biggl \}\, .
\end{equation}
In addition, the perturbative K\"ahler potential \eqref{eq:TreeLevKP} is independent on the axions ${\rm Re}(z^i)$, $i=1,\dots,h^{1,2}$. This results in a discrete $\text{Sp}(2h^{1,2} +2, \mathbb{Z})$ gauge symmetry generating integer shifts of the complex structure moduli
\begin{equation}\label{Ushift}
z^i \to z^i+n^i\,,\quad n^i\in\mathbb{Z}\, ,\quad i=1,\dots,h^{1,2}\,.
\end{equation}
The period vector and the fluxes transform under monodromy as
\begin{equation}
\{\Pi,h,f\} \to M_{\{n^i\}}\{\Pi,h,f\}\,,\qquad M_{\{n^i\}}\in Sp(2h^{1,2} +2, \mathbb{Z})\, .
\end{equation}
These transformations leave the K\"ahler potential \eqref{eq:TreeLevKP}, the superpotential \eqref{eq:superpot},  and the tadpole \eqref{eq:fluxtadpole} invariant. By using these integer shifts in Eq.~\eqref{Ushift}, we can choose the fundamental domain for the axions as ${\rm Re}(z^i)\in (-0.5,0.5]$.

Complex structure moduli stabilisation is the process of identifying minima of the flux-induced scalar potential \eqref{eq:scalarpotential}. In this work we focus on flux vacua satisfying the $F$-flatness conditions
\begin{subequations}
\label{eq:fflat}
\begin{align}
  D_{\tau}  W&=\frac{1}{\overline{\tau}-\tau}(f-\overline{\tau}h)^{T}\cdot \Sigma\cdot \Pi(z)=0\, , \label{eq:fflat1}\\
  D_i  W&= (f-\tau h)^{T}\cdot \Sigma\cdot\left(\partial_{i} \Pi(z)+\Pi(z)\partial_{i} K\right)=0\, . \label{eq:fflat2}
\end{align}
\end{subequations}
For later purposes, we note that these conditions are equivalent to the imaginary self-duality (ISD) of $3$-form $G_3$, i.e., $\star_6  G_{3}=\I  G_{3}$ in terms of the Hodge star operator $\star_6$ on $X_3$~\cite{Giddings:2001yu}. In terms of the flux vectors \eqref{eq:fluxdef}, it can be written as
\begin{equation}
\label{eq:ISD_complex}
    f_1-\tau\, h_1 = \overline{\vphantom{A^a}\cN\,} \cdot (f_2-\tau\,  h_2)
\end{equation}
where $\overline{\vphantom{A^a}\cN\,}$ is the (complex conjugate) gauge kinetic matrix defined in terms of the prepotential as
\begin{equation}
\label{eq:GaugeKinMatrix}
    \cN_{I J}=\overline{F}_{IJ}+2\I\, \dfrac{\text{Im}(F_{I L})X^{L} \, \text{Im}(F_{J K})X^{K}}{X^{M}\text{Im}(F_{MN})X^{N}}\kom F_{IJ}=\p_{X^{I}}\p_{X^{J}}F\, .
\end{equation}
Alternatively, by using $\tau=c_{0}+\I s$, we can write this ISD condition form
\begin{equation}
\label{eq:ISDCond_real} 
    f=\left (s\, \Sigma\cdot \cM+c_{0}\mathds{1}\right )\cdot h
\end{equation}
in terms of the real matrix
\begin{equation}
\label{eq:ISD_matrix}
    \cM = \left (\begin{array}{cc}
                     \; -\cI^{-1} &\; \cI^{-1}\cR  \\ 
                       \;\;\cR\cI^{-1} &\,  -\cI-\cR \cI^{-1}\cR\;
                       \end{array} \right )\, ,
\end{equation}
which we refer to as \emph{ISD matrix} subsequently.
Here, $\cR,\cI$ are the real and imaginary parts of the gauge kinetic matrix $\cN=\cR+\I\,\cI$ defined above.

Early attempts to construct vacua solving \eqref{eq:fflat1} and \eqref{eq:fflat2} include \cite{Giryavets:2003vd,Giryavets:2004zr,DeWolfe:2004ns,Denef:2004dm,Conlon:2004ds,Eguchi:2005eh}, see also \cite{Martinez-Pedrera:2012teo,Cicoli:2013cha,Brodie:2015kza,Blanco-Pillado:2020wjn,Blanco-Pillado:2020hbw} for models with $h^{1,2}\leq 3$.\footnote{An alternative strategy is to restrict to special choices of fluxes for which a subset of VEVs can be fixed analytically, see e.g. \cite{Demirtas:2019sip,Marchesano:2021gyv,Coudarchet:2022fcl}.}
The distributions of string vacua have been studied in detail in \cite{Douglas:2003um,Ashok:2003gk,Douglas:2004zg,Denef:2004ze,Douglas:2006zj,Lu:2009aw,Cheng:2019mgz} making use of the continuous flux approximation.
For a given value of $N_{\mathrm{flux}}$, the finiteness of flux vacua satisfying \eqref{eq:fflat1} and \eqref{eq:fflat2} has been proven in \cite{Grimm:2020cda,Bakker:2021uqw}.\footnote{The arguments of \cite{Grimm:2020cda,Bakker:2021uqw} actually concern self-dual classes in F-theory which extend to imaginary self-dual fluxes in the weak coupling limit for Type IIB orientifolds studied in this work.}
Subsequently, the authors of \cite{Plauschinn:2023hjw} developed a constructive procedure for enumerating, at least in principle, all flux vacua in a given Type IIB orientifold compactification. With this, \cite{Plauschinn:2023hjw} computationally confirmed the finiteness of $F$-flat vacua in a simple one-modulus case, namely an orientifold of the mirror octic with $Q_{D3}=8$. Below, we will describe and develop further the ideas presented in \cite{Plauschinn:2023hjw}.
 
Hereby, we make use of a systematic framework for numerically constructing flux vacua that was recently developed by some of the authors in \cite{Dubey:2023dvu} making the regime $h^{1,2}\gtrsim 10$ accessible. As a first application, it has been employed in \cite{Ebelt:2023clh} to collect millions of flux vacua for $20$ different CY orientifold compactifications and compare the distributions of the vacuum expectation value (VEV) $W_0$ of the gauge-invariant\footnote{Let us note that, under $\mathrm{SL}(2,\bZ)$ transformations \eqref{eq:SLtrafo}, the value of $W_0$ only changes by a phase so that $|W_0|$ remains invariant.} GVW-superpotential\footnote{The normalisation is chosen based on the conventions of~\cite{Kachru:2019dvo,Demirtas:2019sip}.}
\begin{equation}
\label{eq:superpot}
    W_0 = \sqrt{\frac{2}{\pi}}\, \biggl\langle \mathrm{e}^{K/2} W\biggl\rangle\, .
\end{equation}
Similarly, supersymmetry breaking vacua with quantised fluxes were obtained in \cite{Krippendorf:2023idy} for which, instead of \eqref{eq:fflat1} and \eqref{eq:fflat2}, the extremum conditions $\p_\tau V=\p_{z^i}V=0$ need to be solved. Once combined with Kähler moduli stabilisation, such solutions can be used for $F$-term uplifting to de Sitter vacua in string theory \cite{Saltman:2004sn}, see also \cite{Gallego:2017dvd} for early attempts in the continuous flux approximation.

\section{Targeted explorations of the string landscape}
\label{sec:3}

In this section we derive bounds on the number of flux vacua satisfying $D_IW=0$ in finite regions $U$ of moduli space for given values of the flux induced D3-charge $N_{\text{flux}}\leq N_{\rm max}$ less than some maximum D3-charge $N_{\rm max}\leq Q_{D3}$. With these at hand, we describe an algorithm to numerically construct all solutions in $U$, at least in principle.

Before we begin, let us motivate the need for more targeted explorations of the flux landscape. For one, the generation of flux vacua from uniformly sampled fluxes is ineffective, see e.g.~\cite{Dubey:2023dvu} for a comparison of different sampling strategies. Even more importantly, at least in compactifications on CY hypersurfaces from the Kreuzer-Skarke list \cite{Kreuzer:2000xy}, it becomes increasingly challenging to land inside the large complex structure region of moduli space for large $h^{1,2}$ \cite{Demirtas:2018akl,Plauschinn:2021hkp} because the K\"ahler cones \eqref{eq:DefKahlerCone} get narrower. This demands a more targeted approach to constructing string vacua than a random search as recently initiated in \cite{Plauschinn:2023hjw}.

\subsection{Bounding the flux landscape}
\label{sec:bounds}

Let us start by deriving bounds on the available flux choices in finite regions of moduli space with $N_{\text{flux}}\leq N_{\rm max}$. Hereby, we mainly follow \cite{Plauschinn:2023hjw}, but we add new bounds on the choices of flux vectors $f,h\in \bZ^{2(h^{1,2}+1)}$ entering the GVW superpotential \eqref{eq:SupPotPerVec}. Specifically, we want to bound the fluxes for ISD sampling \cite{Dubey:2023dvu}. The basic idea is to fix points in moduli space together with a subset of flux quanta, and fix the remaining fluxes through the ISD condition \eqref{eq:ISD_complex} or alternatively \eqref{eq:ISDCond_real} \cite{Denef:2004dm,Tsagkaris:2022apo,Dubey:2023dvu}. As will be explained in the subsequent subsection, we use these constraints to devise algorithms to collect all fluxes for given $N_{\rm max}$.

Initially, let us define $U\subset \cM_{\text{cs}}(X_3)\times \cM_\tau$ as an open neighbourhood in complex structure moduli space and the fundamental domain $\cM_\tau$ of the axio-dilaton $\tau=c_0+\I s$ as defined in Eq.~\eqref{eq:FDAD}.
Next, we note that the ISD-matrix $\cM$ defined in \eqref{eq:ISD_matrix} is real, symmetric ($\cM^{T} = \cM $), symplectic ($ \cM^{T} \Sigma \cM   = \Sigma $) and positive definite (i.e. eigenvalues $\lambda_{I}>0$). Most importantly, the real eigenvalues come in pairs satisfying
\begin{equation}
    (\lambda_{I}, \lambda^{-1}_{I}) \qquad \text{with} \qquad\lambda_{I} > 1\, .
\end{equation}
These eigenvalues monotonically increase in the limit of large complex structure.
Basic inequalities for matrix norms suggest that \cite{Plauschinn:2023hjw}
\begin{equation}
\label{eq:hbound}
    \frac{1}{\lambda_{\text{max}}} ||h||^2 \leq  \int H_3 \wedge \star H_3 = h^{T}\cM h \leq \lambda_{\text{max}}||h||^2 \, ,
\end{equation}
where $\lambda_{\text{max}}$ is the maximal eigenvalue of $\cM$.

We solve \eqref{eq:fflat1} explicitly for the axio-dilaton in terms of fluxes and the moduli by writing
\begin{equation}
\label{eq:mataxiodilaton}
    c_0 = \frac{\int H_3\wedge \star F_3}{\int H_3\wedge \star H_3}= \frac{h^T\cM f}{h^T\cM h} \kom s = \frac{N_{\text{flux}}}{\int H_3\wedge \star H_3}= \frac{N_{\text{flux}}}{h^T\cM h}\, .
\end{equation}
Then, by combining \eqref{eq:hbound} with \eqref{eq:mataxiodilaton} and \eqref{eq:FDAD}, one finds that the Euclidean norm of $h$ can be constrained as \cite{Plauschinn:2023hjw}
\begin{equation}
\label{eq:CondHfluxFullVec} 
    ||h||^{2}\leq \dfrac{2N_{\text{max}}\lambda_{\text{max}}}{\sqrt{3}}\, .
\end{equation}
Vice versa, this allows us to bound the dilaton from above as
\begin{equation}
\label{eq:bound_dilaton}
    \frac{\sqrt{3}}{2}\leq s \leq  \lambda_{\text{max}} N_{\text{max}}\, .
\end{equation}
This bound can be further improved by plugging \eqref{eq:ISDCond_real} into \eqref{eq:fluxtadpole} and again using elementary identities for the eigenvalues of $\mathcal{M}$ to arrive at
\begin{equation}
\label{eq:bound_dilaton_strong}
    \dfrac{\sqrt{3}}{2} \leq s \leq \dfrac{\lambda_{\rm max}N_{\rm max}}{||h||^2}+\dfrac{||h||^2}{4\lambda_{\rm max}}\, .
\end{equation}
This is a slightly stronger bound than \eqref{eq:bound_dilaton}, especially for large $||h||^2$.

Next, let us write $\cN=\cR+\I \cI$ and $\cN^{-1}=\tilde{\cR}+\I\tilde{\cI}$.
Then one can show that, by using the ISD condition \eqref{eq:ISD_complex},
the value of the tadpole contribution from fluxes can be written as
\begin{equation}
\label{eq:TadpoleISD}
    N_{\rm flux} = \begin{cases}
                            s\, h_2^T\cdot (-\cI)\cdot h_{2}+\frac{1}{s} (f_2-c_0h_2)^T\cdot (-\cI)\cdot (f_2-c_0h_2)\, , & \\[0.5em]
                            s\, h_{1}^T\cdot \tilde{\cI}\cdot h_{1}+ \frac{1}{s} (f_1-c_0h_1)^T\cdot \tilde{\cI}\cdot (f_1-c_0h_1) \, . &
                       \end{cases}
\end{equation}
Notice that $(-\cI)_{JI}$ and $\tilde{\cI}^{IJ}$ are positive definite. Let us denote the eigenvalues of $(-\cI)$ as $\mu$ and those of $\tilde{\cI}$ as $\tilde{\mu}$.
Then we have
\begin{subequations}
\label{eq:CondHflux12}
\begin{align}
  \dfrac{\sqrt{3}}{2} \mu_{\text{min}} ||h_{2}||^{2} &\leq N_{\rm flux}\, ,\label{eq:CondHflux12_h2}\\
  \dfrac{\sqrt{3}}{2} \tilde{\mu}_{\text{min}} ||h_{1}||^{2} &\leq N_{\rm flux}\, . \label{eq:CondHflux12_h1}
\end{align}
\end{subequations}
We stress that these bounds are stronger than the ones of \cite{Plauschinn:2023hjw} where both right hand sides in \eqref{eq:CondHflux12} are bounded by $N_{\rm flux}\lambda_{\text{max}}$. The condition on the LHS \eqref{eq:CondHflux12_h2} on $h_{2}$ is more restrictive than in \eqref{eq:CondHflux12_h1} because the eigenvalues $\mu$ of $(-\cI)$ are larger than the ones of $\tilde{I}$. We will comment on detailed results in Sec.~\ref{sec:results}.

Similarly, we can use \eqref{eq:TadpoleISD} to obtain bounds on the choices of RR-flux vectors $f_1,f_2$ by writing
\begin{subequations}
\label{eq:CondFflux12}
\begin{align}
  \mu_{\text{min}} ||f_{2}-c_0h_2||^{2} \leq\mu_{\text{min}} \biggl (||f_{2}-c_0h_2||^{2}+\dfrac{3}{4}  ||h_{2}||^{2}\biggl ) &\leq s N_{\rm flux} \leq \lambda_{\text{max}} N_{\text{max}}^2 \, ,\label{eq:CondFflux12_f2}\\
   \tilde{\mu}_{\text{min}} ||f_{1}-c_0 h_1||^{2} \leq\tilde{\mu}_{\text{min}} \biggl (||f_{1}-c_0 h_1||^{2}+\dfrac{3}{4} ||h_{1}||^{2}\biggl ) &\leq s N_{\rm flux} \leq \lambda_{\text{max}} N_{\text{max}}^2\, . \label{eq:CondFflux12_f1}
\end{align}
\end{subequations}
The right hand side of both inequalities is less constraining than the ones in \eqref{eq:CondHflux12} due to an extra factor of $N_{\text{max}}$. Thus, we typically expect to find more independent flux choices of $f_1,f_2$ than $h_1,h_2$.
Since the left hand side of \eqref{eq:CondFflux12} involves the value of the universal axion $c_0=\text{Re}(\tau)$, we can relax the above bounds by expanding the terms in \eqref{eq:TadpoleISD} first and then using bounds on the matrix norms to arrive at
\begin{subequations}
\label{eq:CondFflux12Mod}
\begin{align}
  \mu_{\text{min}} \left [||f_{2}||^{2}+\biggl (c_0^2+\frac{3}{4}\biggl )||h_2||^{2}\right ]-2\mu_{\text{max}}\,|c_0|\, ||f_{2}||\, ||h_2|| &\leq sN_{\rm flux}\, ,\label{eq:CondFflux12_f2Mod}\\
    \tilde{\mu}_{\text{min}} \left [||f_{1}||^{2}+\biggl (c_0^2+\frac{3}{4}\biggl )||h_1||^{2}\right ]-2\tilde{\mu}_{\text{max}} \,|c_0|\, ||f_{1}||\, ||h_1|| &\leq sN_{\rm flux}\, . \label{eq:CondFflux12_f1Mod}
\end{align}
\end{subequations}
This then allows us to derive the weaker bounds upon using $|c_0|\leq 0.5$
which are then independent of $c_0$. Later on, these bounds will serve as useful consistency checks for the datasets discussed in Sec.~\ref{sec:results}.

Lastly, there are additional bounds on the RR-fluxes $f$ such as \cite{Plauschinn:2023hjw}
\begin{equation}
\label{eq:fnorm}
    ||f||^2 \leq \frac{4 N^2_{\text{max}}\lambda^2_{\text{max}}}{3}\, .
\end{equation}
The right hand side is much weaker compared to the bound \eqref{eq:CondHfluxFullVec} on $h$. The bound \eqref{eq:fnorm} can be slightly improved by using \eqref{eq:bound_dilaton_strong} to arrive at \cite{Plauschinn:2023hjw}
\begin{equation}
\label{eq:fh_constraint}
   \dfrac{\sqrt{3}}{2}\,\dfrac{N_{\mathrm{flux}}}{\lambda_{\mathrm{max}}} \leq ||f||^2 \leq \dfrac{\lambda_{\rm max}^2  N_{\rm flux}^2}{||h||^2}+\dfrac{||h||^2}{4}\, .
\end{equation}

\subsection{Algorithms for finding flux vacua}
\label{sec:algorithm}

\begin{algorithm}[t!]
\begin{algorithmic}[1]
\Procedure{Flux vacua generation}{$N_{\mathrm{max}},U$}
    \State Generate sample $\mathcal{S}\subset U$
    \State Compute eigenvalues $\mu_{\text{max}}$, $\tilde{\mu}_{\text{max}}$, $\lambda_{\text{max}}$ for all $(z^i,\tau)\in\mathcal{S}$
    \State Generate $h_1$, $h_2$ fluxes subject to \eqref{eq:CondHflux12} in box $\mathcal{H}\in \bZ^{h^{1,2}+1}$ of size set by \eqref{eq:CondHfluxFullVec}
    \For{$h_2\in\mathcal{H}$ \& $h_1\in\mathcal{H}$}
        \If{\eqref{eq:BoundH1H2Comb} $=$ \texttt{True}}
            \State Define $h=(h_1,h_2)$
            \For{$(z_0^i,\tau_0)\in\mathcal{S}$}
                \State Compute $\tilde{f}\in \bR^{2(h^{1,2}+1)}$ from \eqref{eq:ISDCond_real}
                \State Round $\tilde{f}$ to quantised fluxes $f\in \bZ^{2(h^{1,2}+1)}$
                \If{\eqref{eq:fh_constraint} $=$ \texttt{True} \& $N_{\text{flux}} \leq N_{\text{max}}$}
                    \State Solve \eqref{eq:fflat} for $(f,h)$ with initial guess $(z_0^i,\tau_0)$
                    \State Apply $\mathrm{SL}(2,\bZ)$ and $\mathrm{Sp}(2h^{1,2}+2,\bZ)$ transformations
                    \If{$(\langle z^i\rangle, \langle \tau\rangle)\in U$}
                        \State Return $(f,h,\langle z^i\rangle, \langle \tau\rangle)$
                    \EndIf
                \EndIf
            \EndFor
        \EndIf
    \EndFor
\EndProcedure
\end{algorithmic}
\caption{Algorithm to generate flux vacua for given $N_{\mathrm{max}},U$}
\label{alg:1}
\end{algorithm}

We now detail the algorithm for numerically generating fluxes and their associated vacua. The goal is to systematically construct minima of the flux scalar potential \eqref{eq:scalarpotential}, satisfying $D_IW=0$, for a given maximal D3-charge $N_{\mathrm{max}}$ and a finite region $U\subset \cM_{\text{cs}}(X_3)\times \cM_\tau$ in complex structure moduli space.
  
To begin, a sample of points $(z^i,\tau)\in \mathcal{S}\subset U$ is generated uniformly within the desired region $U$. The size of this sample may need adjustment based on the size of the region and the desired precision. When attempting to enumerate all flux vacua for a fixed $N_{\text{max}}$, these steps should be repeated across multiple samples until the number of solutions stabilises. At each sampled point, the matrices $\mathcal{M}$ and $\mathcal{N}$, defined in \eqref{eq:ISD_matrix} and \eqref{eq:GaugeKinMatrix} respectively, are evaluated. The global maximum eigenvalues $\mu_{\text{max}}$, $\tilde{\mu}_{\text{max}}$, and $\lambda_{\text{max}}$ are then computed, providing constraints on the fluxes as discussed in Sec.~\ref{sec:bounds}.

Possible choices for $h_1$ and $h_2$ are generated next, ensuring that they satisfy the conditions in \eqref{eq:CondHflux12}. 
To combine them into $h=(h_1,h_2)^T$, we use \eqref{eq:CondHfluxFullVec} to write\footnote{We note that using \eqref{eq:CondHflux12} leads to a weaker bound since $\lambda_{\text{max}}<\mu^{-1}_{\text{min}}+\tilde{\mu}^{-1}_{\text{min}}$.}
\begin{equation}
\label{eq:BoundH1H2Comb}
    ||h_{1}||^{2}\leq\dfrac{2 N_{\text{flux}}}{\sqrt{3}} \lambda_{\rm max}-||h_{2}||^{2}
\end{equation}
That is, for given $h_2$, only those choices of $h_1$ lead to a consistent choice of $h$ for which \eqref{eq:BoundH1H2Comb} is satisfied.
Then, for each valid $h$, the ISD condition \eqref{eq:ISDCond_real} is used to calculate the RR fluxes $\tilde{f}\in\bR^{2(h^{1,2}+1)}$
for each point in $(z_0^i,\tau_0)\in\mathcal{S}$. These fluxes are subsequently rounded to integer values $f\in\bZ^{2(h^{1,2}+1)}$ shifting the true solution away from the original point in $(z_0^i,\tau_0)\in\mathcal{S}$. Pairs $(f,h)$ are retained only if they satisfy both the flux constraint $N_{\text{flux}} \leq N_{\text{max}}$ and \eqref{eq:fh_constraint}. 

Using numerical optimisers (e.g.~from \texttt{scikit-learn}~\cite{scikit-learn}), the $F$-term conditions \eqref{eq:fflat} are solved for each valid pair $(f,h)$ starting with the initial guesses $(z_0^i,\tau_0)$ from the sample $\mathcal{S}$. If necessary, we apply suitable $\mathrm{SL}(2,\bZ)$ and $\mathrm{Sp}(2h^{1,2}+2,\bZ)$ transformations on the output ensuring that $\tau$ and $\mathrm{Re}(z^i)$ are mapped to their respective fundamental domains as described in Sec.~\ref{sec:flux_vacua}. Given a particular solution, we check that the vacuum expectation values $\langle z^i\rangle, \langle \tau\rangle$ lie within the target region $U$. This step is necessary because rounding the RR-fluxes $\tilde{f}\in\bR^{2(h^{1,2}+1)}$ to integer vectors $f\in\bZ^{2(h^{1,2}+1)}$ typically shifts the solution to $D_IW=0$ away from the initial guesses, possibly placing it outside $U$. Finally, duplicate solutions are removed to retain only unique vacua. 

The algorithm, as summarised in Algorithm~\ref{alg:1}, enables systematic enumeration of all flux vacua within the constraints. It is particularly effective when the values for $N_{\text{max}}$, the region $U$ and the sample $\mathcal{S}\subset U$ are chosen judiciously. While a full classification of all solutions is computationally infeasible for orientifolds with large $h^{1,2}$ or large $Q_{D3}$, the method is well-suited for targeted explorations in selected regions of moduli space.

Several enhancements can improve the computational efficiency of the algorithm. For example, certain \texttt{for} loops can be vectorised using tools such as \texttt{jax.vmap}, and intermediate results, such as the choices of RR-fluxes $f$, can be discarded from memory after use to reduce resource requirements. Additionally, linear approximations of shifts $(\delta z^i, \delta \tau)$ away from initial guesses $(z_0^i,\tau_0)$ can be used to estimate solutions more accurately \cite{Schachner:ISD2024}. More specifically, expanding the ISD condition \eqref{eq:ISDCond_real} to linear order in $\delta f = f-\tilde{f}$ and $(\delta z^i, \delta \tau)$ allows for an analytical solution of the linear system, which provides significantly better estimates of the true $F$-term solutions. If these improved estimates place the solution outside $U$, the corresponding flux pair $(f,h)$ can be discarded without performing a full numerical minimisation, thereby avoiding the most expensive step of the algorithm. As will be shown in \cite{Schachner:ISD2024}, such strategies significantly enhance the optimiser’s efficiency, especially for larger values of $N_{\text{max}}$.

\section{Application at $h^{1,2}=2$}
\label{sec:results}

Let us now turn to finding explicit flux vacua. Our discussion has been quite general until now. To find explicit vacua, we will focus on a particular Calabi-Yau orientifold, namely a $\mathbb{Z}_2$-involution of the degree-18 hypersurface in $\mathbb{CP}^4_{[1,1,1,6,9]}$, see e.g. \cite{Candelas:1994hw, Martinez-Pedrera:2012teo, Demirtas_2020} for previous studies. This has $h^{1,2} =272$ and $h^{1,1}=2$. As in \cite{Giryavets:2003vd}, we will work at the symmetric locus of its $\Gamma = {\mathbb{Z}}_{6} \times {\mathbb{Z}}_{18}$ discrete symmetry. This reduces the effective number of complex structure moduli to two and significantly reduces the computational complexity involved in finding explicit vacua.

The defining equation for this hypersurface is given by
\begin{equation}
x_1^{18}+ x_2^{18}+ x_3^{18}+x_4^{18}+x_5^{18}-18\psi x_1x_2x_3x_4x_5 -3\phi x_1^6x_2^6x_3^6=0\,,
\end{equation} 
where $\psi$ and $\phi$ are the complex structure deformations invariant under $\Gamma$. As discussed in Sec.~\ref{sec:lcs}, the leading term in the prepotential is given by the intersection numbers of the mirror dual $\widetilde{X}_3$. For the effective theory of the two moduli, these are
\begin{equation}
    \widetilde{\kappa}_{111}=9\, , \ \widetilde{\kappa}_{112}=3, \ \widetilde{\kappa}_{122}=3\, ,  \ b=\frac{1}{2} \begin{pmatrix} 9 & 3 \\ 3 & 0 \end{pmatrix}\, , \ a=\frac{1}{4}\begin{pmatrix}
        17 \\ 6
    \end{pmatrix}\, .
\end{equation}
The first and second instanton corrections to the prepotential are given by
\begin{align}
    (2\I\pi)^3{F}_{\text{inst}} &=-540q_1 -3q_2-\frac{1215}{2}q_1^2 +1080q_1q_2 +\frac{45}{8}q^2_2+\dots\, ,
\end{align}
where $ q_i = \mathrm{e}^{2\pi \I z^i}$. We will construct vacua in the regime where the instanton corrections can be safely ignored. In practice, we find that in our solutions
\begin{equation}
\label{eq:inst_suppression}
    \frac{|F_{\text{inst}}|}{|F|}, \  \frac{540 \mathrm{e}^{-2\pi \text{Im}(z^1)}}{(2\pi)^3|F|}, \ \frac{3 \mathrm{e}^{-2\pi \text{Im}(z^2)}}{(2\pi)^3|F|}\ \leq \ 10^{-5} \, .
\end{equation}
Below, we present our results for flux vacua obtained for specific regions in moduli space by using the algorithm described in Sec.~\ref{sec:algorithm} and the bounds of Sec.~\ref{sec:bounds}. Specifically, we systematically construct flux vacua for $N_{\text{flux}}\leq N_{\text{max}}$ with $N_{\text{max}}=50$ in regions contained in
\begin{equation}
\label{eq:tregion}
    U \subset \biggl \{\mathrm{Re}(z^i)\in (-0.5,0.5]\, ,\; \mathrm{Im}(z^i)\in [1,10]\, ,\;
    c_0\in (-0.5,0.5]\, ,\; s\in \left [\frac{\sqrt{3}}{2},50\right ] \biggl \}\, .
\end{equation}
We stress again that we collect only gauge inequivalent vacua under $\text{Sp}(6,\mathbb{Z})$ and $\text{SL}(2,\mathbb{Z})$ gauge symmetries, recall the discussion in Sec.~\ref{sec:flux_vacua}. For our example, the monodromy shifts under $\text{Sp}(6,\mathbb{Z})$ are generated by matrices $M_{\{n_1,n_2\}}$ computed in e.g. \cite{Cicoli:2022vny}.

\subsection{Numerical ensembles}

\begin{table}[t!]
    \centering
    \resizebox{1.\textwidth}{!}{
    \begin{tabular}{| c || c | c | c || c | c | c | c || c |}
         \hline
         & & & & & & & & \\[-1.3em]
         Name & $\text{Im}(z^i)$ & $s$ & $N_{\text{max}}$ & \#$h$ & \#$f$ & \#$(f,h)$ & $\mathcal{N}_{\mathrm{vac}}$ & exhaustive\footnotemark \\[0.2em]
         \hline
         \hline 
         & & & & & & & & \\[-1.2em]
         A & $[2,3]$ & $\bigl [\frac{\sqrt{3}}{2},20\bigl ]$ & $34$ & 82,082 & 1,849,426 & 5,134,862 & 5,140,872 & Yes \\[0.3em]
         \hline
         & & & & & & & & \\[-1.2em]
         B & $[2,5]$ & $\bigl [\frac{\sqrt{3}}{2},10\bigl ]$ & $10$ & 1,900 & 6,340 & 12,160 & 12,196  & Yes \\[0.3em]
         \hline
         & & & & & & & & \\[-1.2em]
         C & $[1,10]$ & $\bigl [\frac{\sqrt{3}}{2},50\bigl ]$ & $34$ & 3,652,744 & 21,043,832 & 50,652,686 & 50,884,086 & No \\[0.3em]
         \hline
         & & & & & & & & \\[-1.2em]
         D & $[2,10]$ & $\bigl [\frac{\sqrt{3}}{2},10\bigl ]$ & $50$ & 5,909,012 & 45,886,900 & 123,075,206 & 123,408,240 & No \\[0.3em]
         \hline 
    \end{tabular}
    }
\caption{Summary of flux vacua gathered for different regions $U$ as specified in Eq.~\eqref{eq:tregion}. With \#$h$, \#$f$, \#$(f,h)$, we denote respectively the number of unique NSNS-fluxes, RR-fluxes and full flux choices. Further, the total number of vacua is given by $\mathcal{N}_{\mathrm{vac}}$. The fluxes and moduli VEVs can be found on the following GitHub repository \href{https://github.com/ml4physics/JAXvacua}{\texttt{https://github.com/ml4physics/JAXvacua}}. 
}\label{tab:summary}
\end{table}
\footnotetext{With exhaustive, we mean that running the algorithm of Sec.~\ref{sec:algorithm} for more samples does not give rise to any new solutions.}

Let us start by employing the algorithm from Sec.~\ref{sec:algorithm} for different choices of $N_{\text{max}}$ and regions $U$ as defined in \eqref{eq:tregion}. In Tab.~\ref{tab:summary} we summarise the counts for different choices of $U$ and values for $N_{\text{max}}$. In particular, we state the number of unique choices of NSNS-fluxes $h$, of RR-fluxes $f$, and full flux configurations $(f,h)$ together with the total number of vacua $\mathcal{N}_{\mathrm{vac}}$. 

For datasets A and B, we performed an exhaustive search, i.e. we enumerate all flux vacua consistent with the bounds of Sec.~\ref{sec:bounds}. Here, we choose either a small enough value for $N_{\text{max}}$ (dataset A) or a suitably small region in moduli space (dataset B) to make a classification of all viable vacua. In contrast, datasets C and D are obtained for larger values of $N_{\text{max}}$ and the moduli. We stress that these datasets are far from being random: the vacua were constructed using a targeted approach of Sec.~\ref{sec:algorithm} as opposed to e.g. a random sampling of fluxes from a uniform distribution. Below, we characterise the datasets A and B and subsequently compare them against the statistical expectations. 

\begin{figure}[t!]
\centering
\includegraphics[width=1.0\columnwidth]{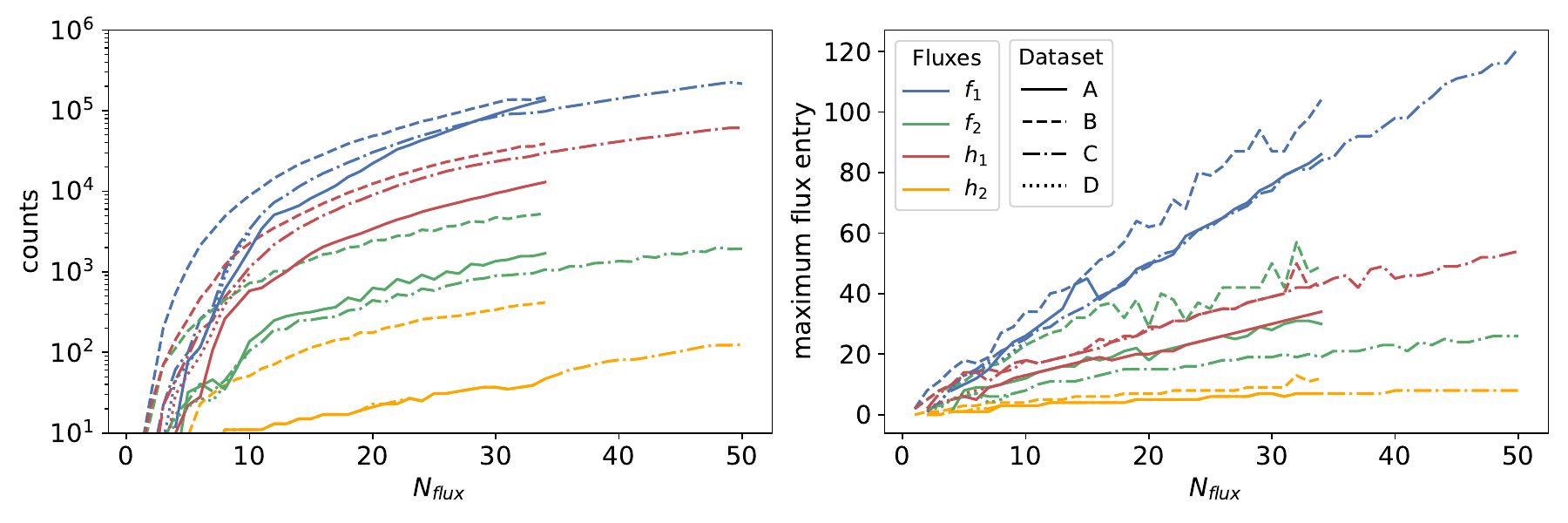}
\caption{Properties of fluxes in our datasets. \emph{Left}: Number of unique choices of flux vectors $f,h\in\bZ^{6}$ and the corresponding unique sub-vectors $f_1,f_2,h_1,h_2\in\bZ^{3}$ (recall Eq.~\eqref{eq:fluxdef}) in our dataset. \emph{Right}: Maximum flux number for each flux vector $f_1,f_2,h_1,h_2\in\bZ^{3}$ bounding the flux lattice in the various directions.}
\label{fig:statflux}
\end{figure}

Initially, we study the features of dataset A. We note that, since the maximum eigenvalue $\lambda_{\mathrm{max}}$ of the ISD matrix $\cM$ is monotonically increasing towards LCS, it becomes increasingly difficult to collect all flux configurations for fixed $N_{\text{max}}$ the larger $\text{Im}(z^i)$. For larger $N_{\text{max}}$, we therefore choose a suitably small region for $\text{Im}(z^i)$ to enumerate all solutions, namely $2\leq \text{Im}(z^i)\leq 3$. Crucially, despite this restriction to a rather small region in moduli space, we find $5,140,872$ flux vacua with $N_{\text{flux}}\leq 34$. This needs to be compared with $15,392$ solutions\footnote{We are quoting here the number of solutions from \cite{Martinez-Pedrera:2012teo} at the large complex structure for which the left-hand side of \eqref{eq:inst_suppression} is less than $10^{-2}$.} obtained in \cite{Martinez-Pedrera:2012teo} of which only a small subset seems to be contained in the region $2\leq \text{Im}(z^i)\leq 3$. Clearly, the algorithm for such a classification of all solutions matters: the authors of \cite{Martinez-Pedrera:2012teo} chose a particular parametrisation of fluxes so that the RR-fluxes $f$ are related to the NSNS-fluxes $h=(h_1,h_2)$ via the relationship $f=(-h_2,h_1)$. This effectively reduces the 12-dimensional flux space to a $6$-dimensional subspace. We can understand quantitatively why this parametrisation is missing a lot of solutions by simply looking at the left plot in Fig.~\ref{fig:statflux}: the unique choices of RR-fluxes $f=(f_1,f_2)$ dominate by roughly one order of magnitude compared to the NSNS-fluxes $h=(h_1,h_2)$. By enforcing $f=(-h_2,h_1)$, the majority of RR-flux choices remain undetected.

Let us point out that one of the advantages of the homotopy continuation method employed in \cite{Martinez-Pedrera:2012teo} is the standard lore that it finds \emph{all} solutions for a given set of input parameters, in our case the fluxes. In fact, they obtained $9.5$ solutions per flux choice, but the majority of them are unphysical. It stands to reason that at first glance our methods provide less guarantees in this regard. We have to keep in mind, however, that we are interested only in solutions in special regions for which our algorithm of Sec.~\ref{sec:algorithm} is perfectly adapted. Overall, we found $5,940$ flux configurations with multiple solutions to the F-flatness equations \eqref{eq:fflat}. 

It is also interesting to contrast the number of consistent flux choices leading to vacua in $U$. In the left panel of Fig.~\ref{fig:statflux}, we present the number of unique integer flux vectors $f,h\in\bZ^{6}$ and the sub-vectors $f_1,f_2,h_1,h_2\in\bZ^{3}$ (recall Eq.~\eqref{eq:fluxdef}). We observe that the counts differ by several orders of magnitude. While $h_2$ seems to be most constrained with only around $\mathcal{O}(10^3)$ different values at $N_{\text{flux}}=34$, we obtained $\mathcal{O}(10^6)$ distinct choices of $f_1$ in our dataset. Let us stress that this behaviour arises due to properties of the ISD condition \eqref{eq:ISD_complex} at large complex structure. The gauge kinetic matrix $\cN$ contains hierarchical entries scaling up the flux entries of $f_1,h_1$ for given $f_2,h_2$. Hence, the former dominates the counting in Fig.~\ref{fig:statflux}.

This reasoning can also be corroborated by examining the individual flux quanta in $f_i$ and $h_i$. The plot on the right in Fig.~\ref{fig:statflux} illustrates the maximum flux entry for each unique flux vector $f_1, f_2, h_1, h_2 \in \mathbb{Z}^3$ in our dataset. This serves as a measure of the largest sphere around the origin in $\mathbb{Z}^3$ that contains all these integer vectors. We observe that the various directions in flux space are bounded differently: while $h_2$ only contains $\mathcal{O}(1)$ values even at large $N_{\text{flux}}$, the maximum entries of $f_1$ reach as high as $\mathcal{O}(80)$. This observation has significant implications for the sampling of fluxes. In many applications, individual subcomponents of $f$ and $h$ are not distinguished. However, as the above indicates, generating all fluxes within a sphere of a given radius around the origin to systematically enumerate solutions for a given tadpole is inefficient. For instance, even at $N_{\text{flux}} = 4$, finding all vacua requires sampling the entries of $f_1$ in the range $[-11, 11]$. These observations motivate further investigation into these constraints for different geometric regions in moduli space. 

\begin{figure}[t!]
\centering
\includegraphics[width=0.85\columnwidth]{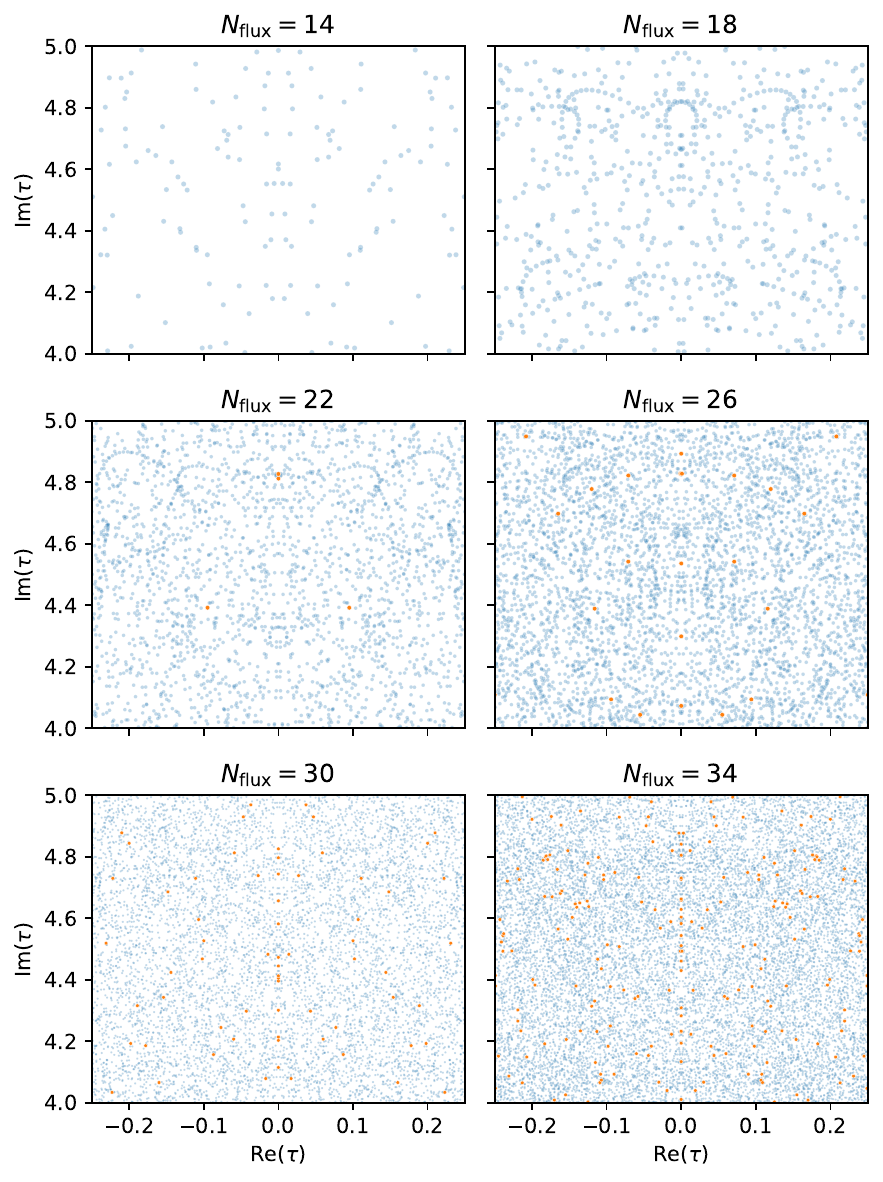}
\caption{Distribution of the axio-dilaton $\tau$ in a small region of its fundamental domain for various values of the flux induced D3-charge $N_{\text{flux}}$ in dataset A. Clusters containing multiple solutions are highlighted in orange.}
\label{fig:tau_struct}
\end{figure}

Let us also comment on the behaviour of structures as we increase $N_{\text{max}}$. In Fig.~\ref{fig:tau_struct}, we show the distribution of the axio-dilaton $\tau$ in a small region of its fundamental domain for various values of the flux induced D3-charge $N_{\text{flux}}$. We highlight clusters with multiple solutions in orange. We clearly observe structural features that are reminiscent of the equivalent plot for the rigid CY as studied e.g. in \cite{Denef:2004ze}. As expected, as the value of $N_{\text{flux}}$ increases, the non-trivial structures are shifted to smaller scales. Here, the intuition is that the relative spacing between the individual vacua scales inversely with $N_{\text{flux}}$. We therefore emphasise that, even in our larger datasets, the distributions exhibit non-trivial patterns that deserve further scrutiny.

\begin{figure}[t!]
\centering
 \includegraphics[width=1.0\columnwidth]{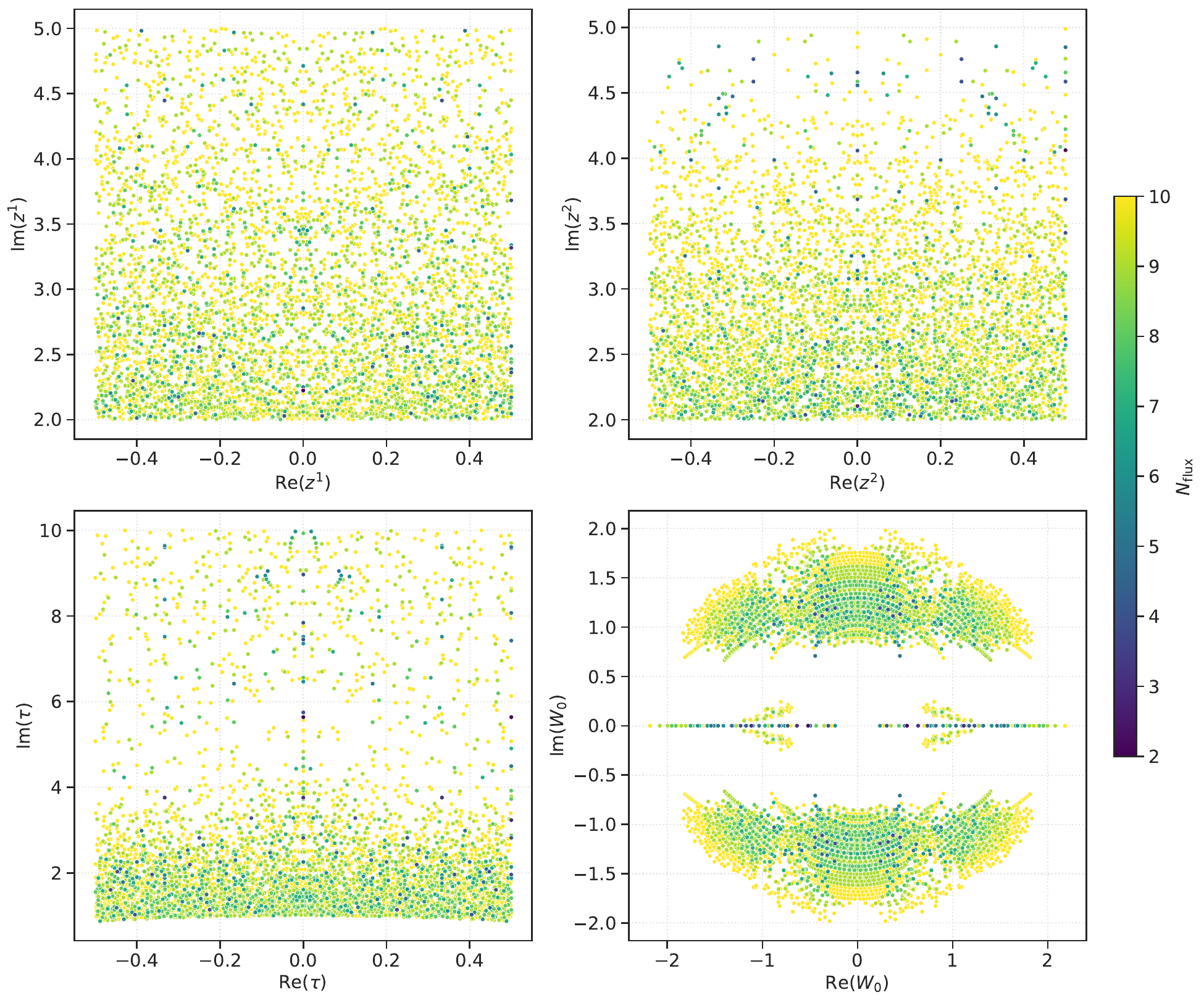}
\caption{Distribution of the complex structure moduli $z^1,z^2$ (top row), the axio-dilaton $\tau=c_0+\I s$ (bottom left) and the superpotential $W_0$ (bottom right) for dataset B.}
\label{fig:moduli_plot}
\end{figure}

Next, we look at dataset B which is smaller in size and hence easier to visualise.
Fig.~\ref{fig:moduli_plot} shows various distributions for dataset B. The top row shows the complex structure moduli $z^1,z^2$ where we clearly observe different structures in each of the distributions. For example, the left plot for $z^1$ contains arc-like structures which cannot be found in the right plot for $z^2$. Let us also note that the symmetry under $\text{Re}(z^i)\rightarrow -\text{Re}(z^i)$ serves as a consistency check of our numerical methods. The distribution of the axio-dilaton $\tau=c_0+\I s$ in the bottom left plot of Fig.~\ref{fig:moduli_plot} exhibits more pronounced structures. In particular, it features clusters, arcs, and voids familiar from earlier work \cite{DeWolfe:2004ns} for the symmetric torus or the rigid CY.\footnote{For an analysis of these topological features using persistent homology, see \cite{Cole:2018emh}.} Again, a useful consistency check for the completeness of our solutions is the symmetry under $\text{Re}(\tau)\rightarrow -\text{Re}(\tau)$.

Lastly, the superpotential $W_0$ is shown in the bottom right plot of Fig.~\ref{fig:moduli_plot}. We recall that our definition for $W_0$ includes the factor of $\mathrm{e}^{K/2}$, cf.~Eq.~\eqref{eq:superpot}. We find that the width of the $W_0$ distribution increases roughly with $\sqrt{N_{\text{flux}}}$ as previously observed in \cite{Ebelt:2023clh}. It is worth pointing out that, since there is a large void near the origin, all solutions lead to moderately large $|W_0|$.

In previous work \cite{Ebelt:2023clh}, two of the authors of this paper studied in detail the distribution of $W_0$ for over twenty different CY orientifolds with varying $h^{1,2}\leq 5$. A common and prominent feature of these results included a highly populated region along the real axis for $\text{Im}(W_0)= 0$. On the one hand, this is actually a gauge artifact: $\mathrm{SL}(2,\bZ)$ transformations can change the phase of $W_0$. Since it is related to a phase only, it is not of significance for most physical observables. On the other hand, the solutions along the real axis turn out to have special properties. We will have more to say about them and the distribution of $W_0$ in Sec.~\ref{sec:W0}.

\subsection{Comparison with statistical expectations}

We now turn to a comparison of the number of vacua $\mathcal{N}_{\text{vac}}$ obtained in our scans with the statistical approach of \cite{Denef:2004ze}. The latter is based on the continuous flux approximation, thereby replacing sums over discrete fluxes by integrals. In this way, it predicts the number of vacua for a given maximum tadpole $N_{\text{max}}$ for $h^{1,2} = 2$ as \cite{Denef:2004ze}
\begin{equation}
\label{eq:ntot}
    \mathcal{N}_{\rm stat}(N_{\rm{flux}}\leq N_{\text{max}})=\frac{(2\pi N_{\text{max}})^{6}}{6!}\int_{\mathcal{M_{\tau}}\times\mathcal{M}_{\text{cs}}} \, \mathrm{d}^{6}z\, \text{det}(g)\, \rho(z)\,. 
\end{equation}
Here, the vacua density $\rho(z)$ in moduli space is given by
\begin{equation}
\label{eq:den}
\rho(z)=\pi^{-6}\int \, \mathrm{d}^2X\, \mathrm{d}^{4}Z \, \mathrm{e}^{-|X|^2-|Z|^2}\, |X|^2\, 
\,\biggl |\text{det}
\begin{pmatrix}
\delta^{IJ}\,\overline{X}-\frac{\overline{Z}^I Z^J}{X} & {F}_{IJK}\,\overline{Z}^K \\
\overline{F}_{IJK}\,{Z}^K & \delta^{IJ}{X}-\frac{Z^I\, \overline{Z}^J}{\overline{X}^{\vphantom{-}}}
\end{pmatrix}\biggl |\, ,
\end{equation}
with $X, Z^1, Z^2$ complex dummy variables.
The model dependence is fully encoded in the rescaled Yukawa couplings $F_{IJK}$ given by
\begin{equation}
 F_{IJK}=-\I\, (e^{\,\,i}_{I})(e^{\,\,j}_{J})(e^{\,\,k}_{K})y_{ijk}\mathrm{e}^{{K}_{\text{cs}}}\kom y_{ijk}=\partial_i\partial_j\partial_k F =  \widetilde{\kappa}_{ijk} + \cO\bigl (\mathrm{e}^{-\text{Im}(z^l)}\bigl )\, ,
\end{equation}
where we are neglecting the terms $\cO\bigl (\mathrm{e}^{-\text{Im}(z^l)}\bigl )$ in the LCS limit. We provide further details on how to compute the integral \eqref{eq:ntot} in App.~\ref{sec:integrations}.

\begin{figure}[t!]
\centering
\includegraphics[width=1.\linewidth]{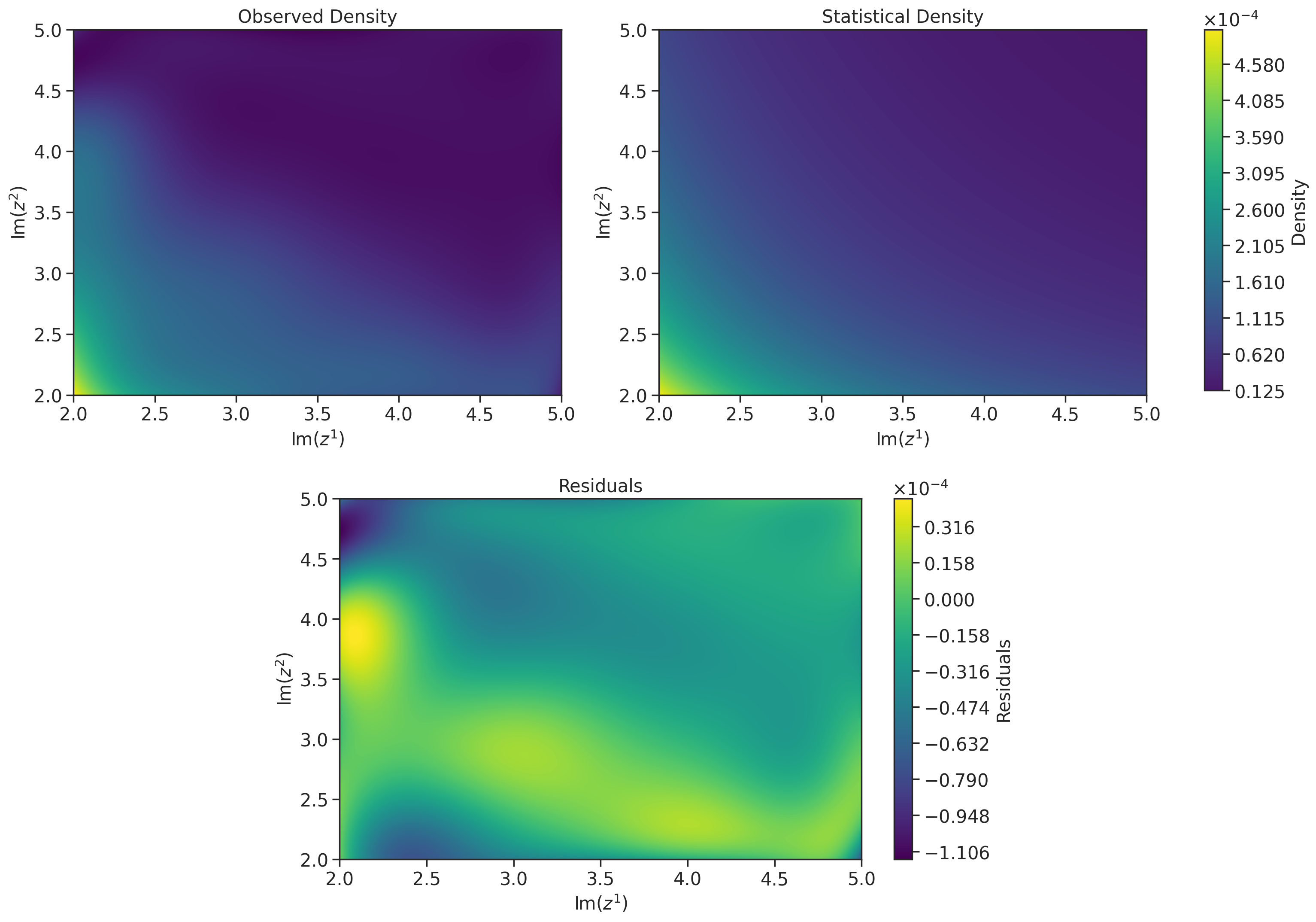}
\caption{\textit{Top}: Comparison of the analytic expectation \eqref{eq:den} for the vacuum density with the observed density found in dataset B. \textit{Bottom}: Difference between the observed and statistical vacuum density.}
\label{fig:den}
\end{figure}

Clearly, as stressed before, a full enumeration of solutions in our example seems to be infeasible. Instead, we compute \eqref{eq:ntot} for certain regions in moduli space and compare the results to our numerical findings for datasets A and B from above. Initially, we are interested in contrasting the vacuum density \eqref{eq:den} with the actual density obtained for dataset A. As evident from Fig.~\ref{fig:den}, our findings reveal significant deviations in some regions: in certain areas of moduli space, we identified more vacua than predicted by the statistical analysis of \cite{Denef:2004ze}, whereas in others, fewer vacua were found. In other words, the actual vacuum density computed in our exhaustive numerical analysis deviates from the analytic expectation \eqref{eq:den} on a local level, as illustrated in Fig.~\ref{fig:den}.

These deviations highlight the importance of combining numerical studies with analytic approaches to gain a more accurate picture of the vacuum distribution. The discrepancies could stem from various factors, including approximations inherent in the continuous fluxes used in \cite{Denef:2004ze}, or the presence of symmetries and degeneracies that are not fully accounted for in the analytic predictions. Furthermore, our results emphasise the role of local moduli space geometry, such as the curvature or clustering of critical points, which may amplify or suppress vacuum densities in specific regions. This interplay between local structure and global expectations suggests that purely statistical treatments may overlook significant variations, motivating a closer examination of local properties in future studies. The deviations shown in Fig.~\ref{fig:den} provide a clear visual representation of these effects, reinforcing the necessity of integrating data-driven methodologies to refine analytic predictions.

\begin{figure}[t!]
\centering
\includegraphics[width=0.8\linewidth]{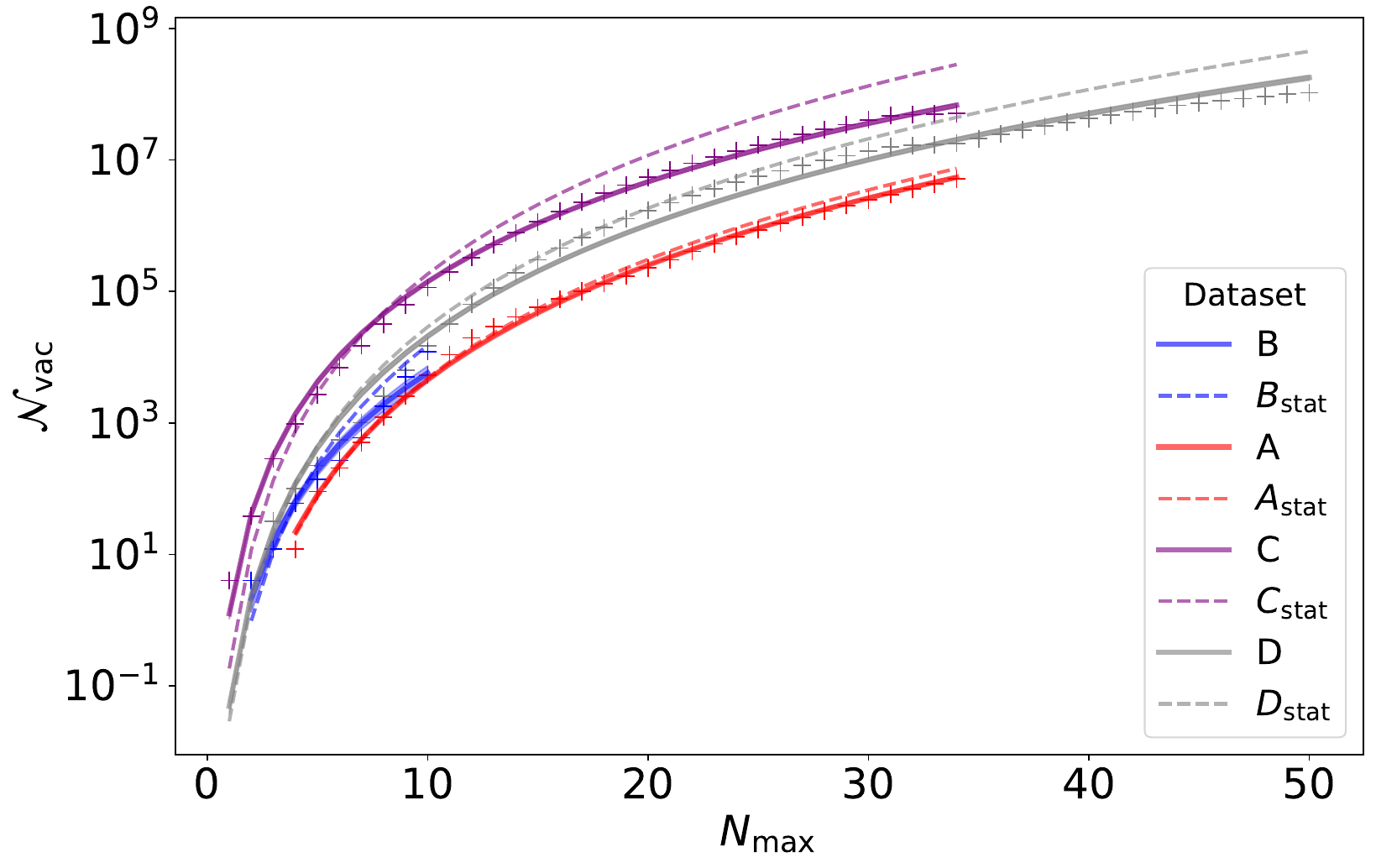}
\caption{Number of vacua with $N_{\rm flux}\leq N_{\text{max}}$ for all datasets from Tab.~\ref{tab:summary}. The crosses are the actual results, the solid curve is the best fit and the dashed curve is the fit for the statistical expected results $\mathcal{N}_{\text{stat}}$.
}\label{fig:N_plot}
\end{figure}

In Fig.~\ref{fig:N_plot} we compare the observed number of vacua $\cN_{\text{vac}}$ with $\cN_{\text{stat}}$ as functions of $N_{\text{max}}$. The best-fit expressions for $\cN_{\text{vac}}$ are
\begin{equation}
\label{eq:vacscale}
    \cN_{\text{vac}} = 
    \begin{cases} 
    (7.17\pm 1.19)\times  10^{-3}\,\times  (N_{\text{max}})^{5.80\pm 0.05} & \text{dataset A}\\[0.4em]
    (7.00\pm 4.19)\times  10^{-2}\,\times  (N_{\text{max}})^{4.91\pm 0.34} & \text{dataset B}\\[0.4em]
    (1.24 \pm 0.20)\, \times (N_{\text{max}})^{5.05\pm 0.06} & \text{dataset C}\\[0.4em]
    (4.89\pm 1.18)\times  10^{-2}\,\times  (N_{\text{max}})^{5.63\pm 0.08} & \text{dataset D}\, .
    \end{cases}
\end{equation}
According to Eq.~\eqref{eq:ntot}, the statistical prediction for the total number of vacua $\cN_{\text{stat}}$ always scales $(N_{\text{max}})^6$. 
More specifically, computing the relevant  integral in \eqref{eq:ntot} for the regions associated with our datasets we find\footnote{Further details are provided in App.~\ref{sec:smallWstat}.}
\begin{equation}
    \cN_{\text{stat}} = (N_{\text{max}})^6\times \begin{cases}
    (0.00483 \pm 1.88 \times 10^{-6}) & \text{dataset A}\\[0.4em]
(0.01535 \pm 7.64 \times 10^{-6}) & \text{dataset B}\\[0.4em]
(0.17458 \pm 8.65 \times 10^{-5}) & \text{dataset C}\\[0.4em]
(0.02886 \pm 1.32 \times 10^{-5}) & \text{dataset D}\, .
    \end{cases} 
\end{equation}

Note that the scalings for the number of vacua ${\mathcal{N}}_{\text vac}$ in \eqref{eq:vacscale} differ from the universal scaling $(N_{\text{max}})^6$ of the statistical predictions. In all cases the observed density scales with a lower power of $N_{\text{max}}$ in comparison with the statistical expectation. A direction for future work is to understand how these deviations depend on the volume of the region of moduli space under consideration and the range of $N_{\rm max}$. More generally, an interesting goal would be to see if there are any global scaling laws for ${\mathcal{N}}_{\text vac}$ across geometries and if so, under which regimes they emerge.

\subsection{IR and UV patterns in our datasets}
\label{sec:pheno}

We now turn to an analysis of the properties of our datasets, some of which relating to their phenomenological aspects (IR properties) and some of which relating to rudimentary imprints of UV-properties, i.e.~how the constraint flux vectors influence our datasets. Our investigation focuses on key properties, including the distribution of $W_0$, the prevalence of vacua with low $|W_0|$, the masses of various moduli, and the hierarchies among them. Understanding these hierarchies is crucial, as they directly influence supersymmetry-breaking scales and the dynamics of low-energy effective field theories. Additionally, we confirm that the lowest value of $|W_0|$ observed in our dataset aligns closely with the predictions of \cite{Denef:2004ze}, providing a quantitative validation of their statistical estimates.

\subsubsection{Distribution of $W_0$}
\label{sec:W0}

\begin{figure}[t!]
\centering
\includegraphics[width=0.75\linewidth]{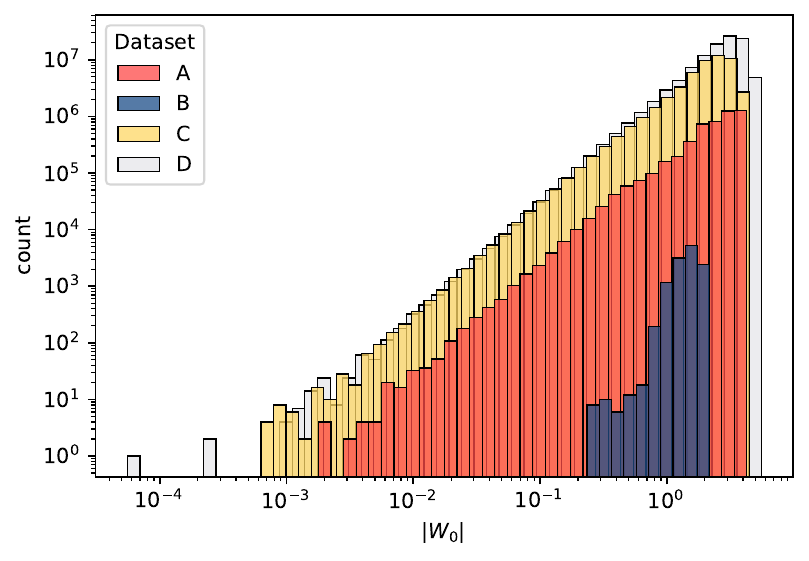}
\caption{Distribution of $|W_0|$ for the four datasets summarised in Tab.~\ref{tab:summary}.}
\label{fig:WGI_hist}
\end{figure}

\begin{figure}[t!]
\centering
\includegraphics[width=\linewidth]{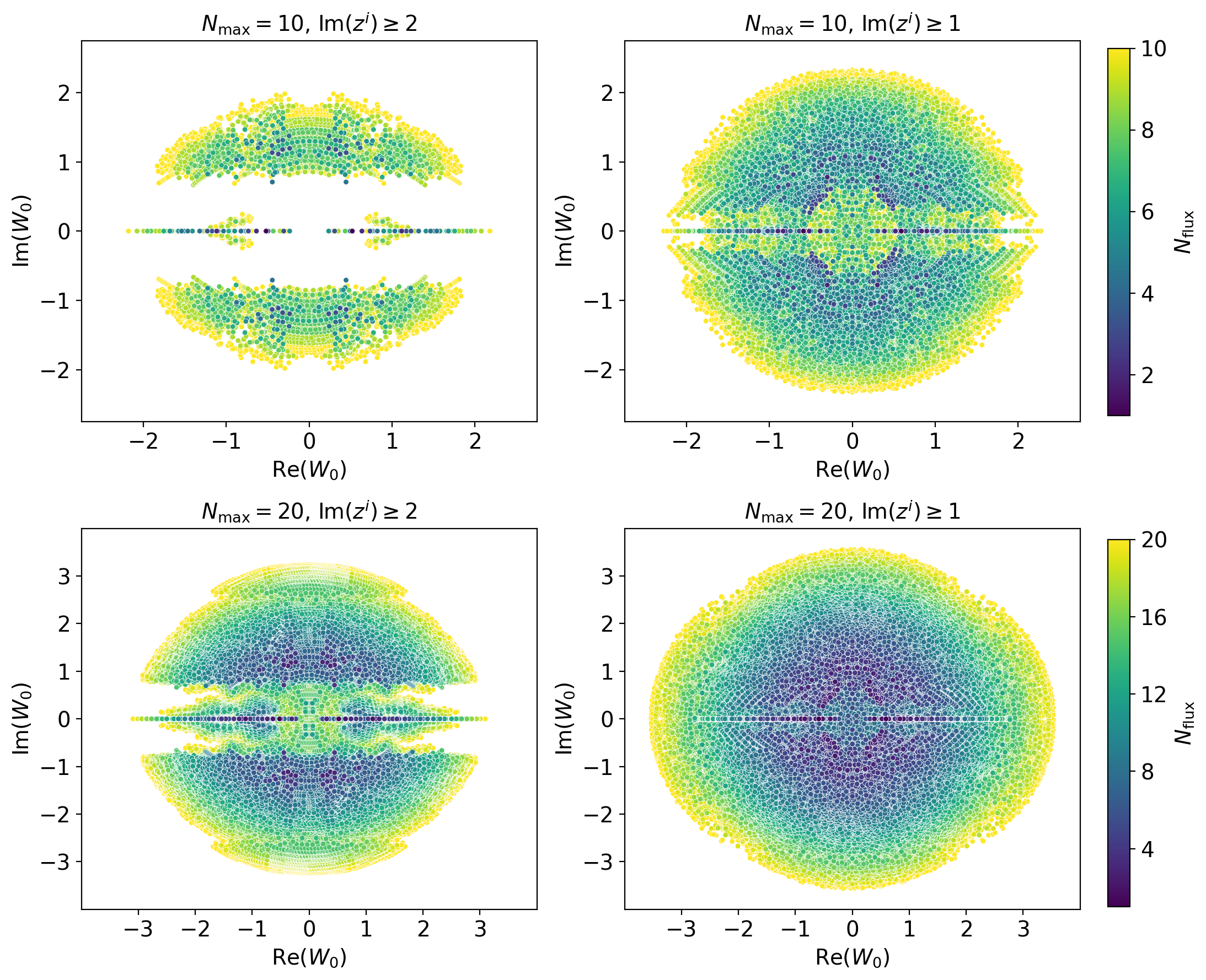}
\caption{\emph{Top}: $W_0$ for vacua with $N_{\text{max}}=10$ for $\text{Im}(z^i) \in [2,5]$ (left) or $\text{Im}(z^i) \in [1,5]$ (right). \emph{Bottom}: $W_0$ for vacua with  $N_{\text{max}}=20$ for $\text{Im}(z^i) \in [2,5]$ (left) or $\text{Im}(z^i) \in [1,5]$ (right).}
\label{fig:WGI_plane}
\end{figure}

We commence our analysis with the distribution of the superpotential $W_0$, depicted in Fig.~\ref{fig:WGI_hist} for the absolute value $|W_0|$, and in Fig.~\ref{fig:WGI_plane} for the corresponding distribution of $W_0$ in the complex plane. For datasets A, C, and D, we observe a universal linear fall-off behaviour of the distribution for $|W_0|$ in Fig.~\ref{fig:WGI_hist} which breaks down at small $|W_0|\lesssim 10^{-2}$. The smallest value of $|W_0|$ in our solutions is
\begin{equation}
    |W_0|=5.547\times 10^{-5}\, ,
\end{equation}
which is obtained from the flux choice
\begin{equation}
    f = (4, 12,  2, -1,  0, -1)\kom h=(36, -1,  0,  0,  1, -1)
\end{equation}
together with the VEVs
\begin{equation}
    \langle z^1\rangle = 0.5 + 2.36817528\, \I\kom \langle z^2\rangle = 0.5 + 2.51175911\, \I\kom \langle \tau\rangle= 0.5 + 1.48121567\, \I\, .
\end{equation}
The small value for $|W_0|$ is here achieved from a purely polynomial superpotential, particularly without having to rely on exponentially small instanton corrections. We stress that this is different to approaches which rely on a perturbatively vanishing $W_0$ and realise a small hierarchy using instanton corrections (see for instance~\cite{Demirtas:2019sip,Broeckel:2021uty}). These solutions do not feature hierarchically suppressed masses and do not rely on instanton corrections.

In Fig.~\ref{fig:WGI_plane} distinctive structural patterns are evident, including a redistribution of points and the emergence of an empty band that disrupts angular symmetry. These features can be partially attributed to the chosen values of the flux quanta, which influence the distribution's overall structure. The observed angular asymmetry arises from gauge fixing, given that the gauge-invariant quantity is the modulus of $W_0$. Nevertheless, the distribution retains symmetry along the $x$- and $y$-axes. Specifically, under reflection about the $x$-axis, the fluxes transform as $(f_2, h_2) \rightarrow (f_2, -h_2)$, while vacua symmetric with respect to the $y$-axis correspond to fluxes related by $(f_2, h_2) \rightarrow (-f_2, h_2)$. These symmetries provide a nuanced understanding of the role fluxes play in shaping the geometry of the vacua distribution.

\begin{figure}[t!]
\centering
  \includegraphics[width=1\linewidth]{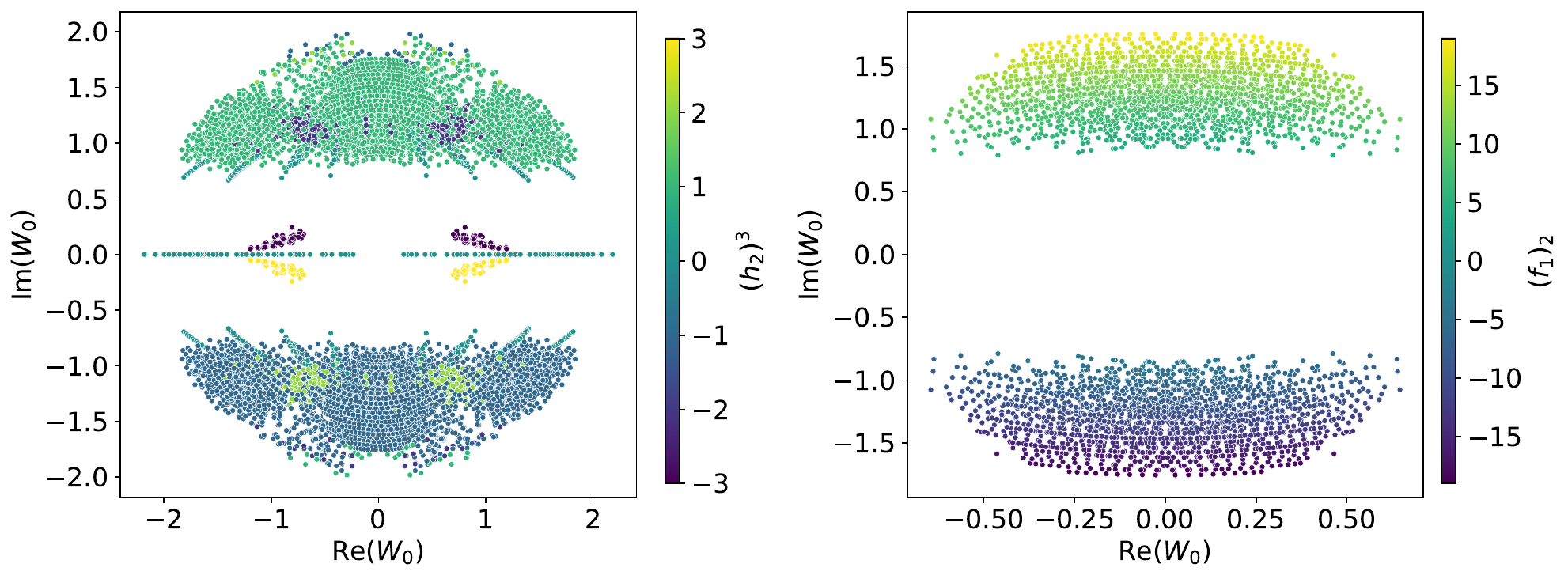}
\caption{\emph{Left:} Distribution of $W_0$ coloured by the flux number $(h_2)^3$ for dataset B. \emph{Right:} Vacua corresponding to fluxes with $|f_2|=(0,1,3)$ and $|h_2|=(0,0,1)$.}
\label{fig:cluster}
\end{figure}

Interestingly, circular arc-shaped structures appear in the distribution, where points with identical values of $|h_2|$ and $|f_2|$ lie on arcs with radii increasing with $|(f_1)_2|$ (see right plot of Fig.~\ref{fig:cluster}). For regions where $\text{Im}(z^i) > 2$, we observe the formation of a predominantly empty band in the range $-0.5 < \text{Im}(W_0) < 0.5$. This empty region is a direct consequence of a hierarchy in the flux quanta $f$ and $h$, induced by the ISD condition \eqref{eq:ISD_complex}, recall the right plot in Fig.~\ref{fig:statflux}. Within this band, isolated clusters emerge, which can be characterised by the value of $(h_2)^3$ (see left plot of Fig.~\ref{fig:cluster}). For points within the band where $\text{Im}(W_0) \approx 0$, the flux quantum $(h_2)^3$ is found to vanish. Analysing the numerical values for the moduli, one finds that the axionic directions $\text{Re}(z^i)$ and $c_0$ always take on rational values. This highlights the intricate interplay between flux quanta, moduli values, and gauge invariance, which collectively shape the observed superpotential distribution.

A noteworthy observation is that there are two limits in which the previously empty band refills. First, including points closer to the boundary of the K\"ahler cone causes the void to almost disappear. This is because the aforementioned flux hierarchies become less dominant for the vacua associated with these points (see the bottom panel of Fig.~\ref{fig:WGI_plane}). Secondly, the spacing between the individual becomes smaller the larger $N_{\text{max}}$, thereby filling in the empty gap in the centre. This also suggests that the minimal value of $|W_0|$ should decrease when scaling up $N_{\text{max}}$ while keeping the region in moduli space fixed. This is of course rather expected according to the statistical analysis of \cite{Denef:2004ze} which we discuss below. Here, we emphasise that small values of $|W_0|$ are indeed available in the interior of the LCS patch of the moduli space provided that $N_{\text{flux}}$ can be tuned large. In many applications, this is precisely required for tadpole cancellation where $N_{\text{flux}}$ needs to be close to the maximally allowed value $Q_{D3}$, recall Eq.~\eqref{eq:fluxtadpole}.

The statistical framework developed in \cite{Denef:2004ze} also offers predictions for the vacua with small $|W_0|$. Specifically, the number of vacua satisfying $|W_0|^2 \leq \lambda_* \ll 1$ and fluxes constrained by $N_{\rm flux} \leq N_{\text{max}}$, for $h^{1,2} = 2$, is given by the expression
\begin{equation}
\label{eq:nsup}
    \cN(\lambda_*, N_{\text{max}}) = \frac{2\pi^4(2N_{\text{max}})^5}{5!}\lambda_* \, I \, , \quad I =  \int_{\mathcal{M}} \dif^6z\, \text{det}(g)\, \mathrm{e}^{2K} F_{ijk} \overline{F}^{ijk}\, ,
\end{equation}
where $\mathcal{M} = \cM_{\text{cs}}\times\cM_{\tau}$ denotes the combined axio-dilaton and complex structure moduli space, and ${F}_{ijk} = \partial_{i}\partial_{j}\partial_{k}F$. This integral encapsulates the complex geometry of the moduli space, accounting for contributions from both the metric determinant $\text{det}(g)$ and $F_{ijk}\overline{F}^{ijk}$.

\begin{table}[t!]
    \centering
    \resizebox{8cm}{!}{
    \begin{tabular}{|c||c|c|}
         \hline
         & & \\[-1.3em]
         Dataset & $\text{min}(|W_0|)_{\text{obs}}$ & $\text{min}(|W_0|)_{\text{stat}}$ \\[0.1em]
         \hline
         \hline
         & & \\[-1.15em]
         A & $1.789\times 10^{-3}$ & $2.192 \times 10^{-3}$ \\[0.15em]
         \hline
         & & \\[-1.15em]
         B & $2.354\times 10^{-1}$ & $2.546 \times 10^{-2}$ \\[0.15em]
         \hline
         & & \\[-1.15em]
         C & $6.305\times 10^{-4}$ & $3.872 \times 10^{-4}$ \\[0.15em]
         \hline
         & & \\[-1.15em]
         D & $5.547\times 10^{-5}$ & $3.324 \times 10^{-4}$ \\[0.15em]
         \hline
    \end{tabular}
    }
    \caption{Observed and statistically predicted minimum $|W_0|$ for datasets summarised in Tab. \ref{tab:summary}.}
\label{tab:nsup}
\end{table}

The minimum achievable superpotential vacua for a given tadpole constraint $N_{\rm{flux}} \leq N_{\text{max}}$ can be estimated by inverting \eqref{eq:nsup} for $\mathcal{N} = 1$, resulting in
\begin{equation}
\label{eq:rootnsup}
  \lambda_* = \frac{5!}{2\pi^4(2N_{\text{max}})^5}\frac{1}{I}\, .
\end{equation}
The integral in \eqref{eq:nsup} can be computed numerically\footnote{We provide further details in App.~\ref{sec:smallWstat}.} using Monte-Carlo methods. Tab.~\ref{tab:nsup} contains the statistically predicted and observed minimum $|W_0|$ values for datasets described earlier in Tab. \ref{tab:summary}. For datasets A and C these two values are quite close; however, for datasets B and D there is a mismatch. Notably, the minimum $|W_0|$ observed in our datasets $5.547\times 10^{-5}$ is an order of magnitude smaller from the smallest $|W_0|$ predicted by statistical analysis, $3.234 \times 10^{-4}$.  

This analysis highlights the interplay between flux quanta, gauge fixing, and the underlying moduli space geometry. The relatively small impact of instanton corrections suggests that the perturbative calculations capture the essential features of the distribution. Moreover, the symmetry considerations elucidate how specific flux transformations influence the localisation and spread of vacua in the moduli space. These insights pave the way for further investigations into the statistical landscape of vacua, particularly in the context of fine-tuning scenarios or additional constraints on $W_0$.

\subsubsection{Moduli masses}

\begin{figure}[t!]
\centering
  \includegraphics[width=1\linewidth]{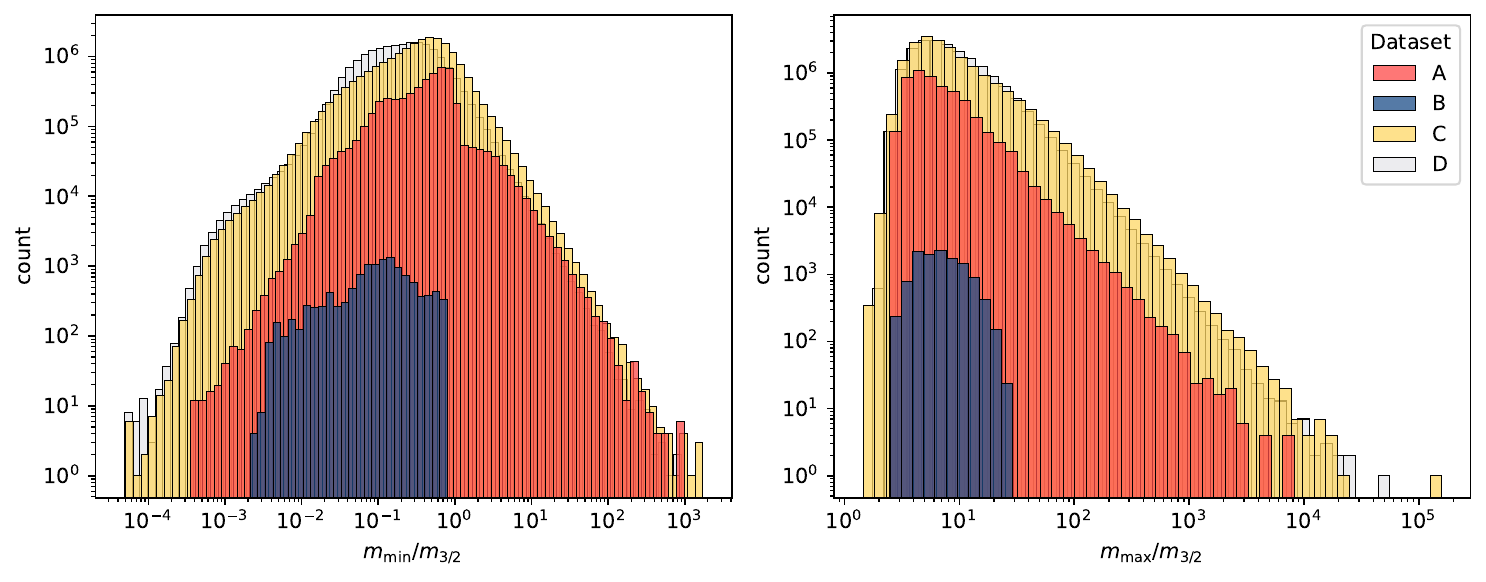}
\caption{Distributions of the maximum and minimum masses for the vacua in each dataset. For datasets C and D, we use subsamples of size $2\times 10^7$.}
\label{fig:mass_dist_minmax}
\end{figure}

Next, we turn to the masses acquired by the moduli fields. First, let us briefly describe the procedure and our results for obtaining masses for the moduli fields $\phi^{I} \subset \{\tau, z^1, z^2\}$. The relevant terms in the $\cN = 1$ supergravity Lagrangian are of the form 
\begin{equation*}
    \mathcal{L} \supset  -K_{I \bar{J}} \partial_{\mu} \phi^{I} \partial^{\mu} \overline{\phi}^{\bar{J}} - V(\phi^{I}, \overline{\phi}^{\bar{J}}),
\end{equation*}
in terms of the scalar potential $V(\phi^{I}, \overline{\phi}^{\bar{J}})$ and the K\"ahler metric $K_{I\Bar{J}} = \partial_{I}\partial_{\bar{J}}K$. The Hessian matrix is given by
\begin{equation}
\label{eq:hessian}
H \equiv \frac{1}{2}
    \begin{pmatrix}
   \frac{\partial^2 V}{\partial \phi^I \partial \phi^J}&\frac{\partial^2 V}{\partial \phi^{I} \partial \overline{\phi}^{\bar{J}}}  \\
     \frac{\partial^2 V}{\partial \overline{\phi}^{\bar{I}} \partial \phi^{J}}& \frac{\partial^2 V}{\partial \overline{\phi}^{\bar{I}} \partial \overline{\phi}^{\bar{J}}}\\
    \end{pmatrix}\, .
\end{equation}
After canonical normalisation of these fields, the eigenvalues of the Hessian matrix \eqref{eq:hessian} provide their squared masses. Since the K\"ahler moduli directions are flat, there is a non-trivial unfixed volume factor ignored in the no-scale scalar potential \eqref{eq:scalarpotential} arising from $\mathrm{e}^{K}$. Owing to this unfixed overall volume normalisation factor, we only look at ratios of the moduli masses. 
Fig.~\ref{fig:mass_dist_minmax} depicts the hierarchical distribution of both maximal and minimal moduli masses with respect to the gravitino mass. The minimum masses range from approximately $10^{-4}$ to $10^{3}$, while the maximum masses are of the order of $1$ to $10^{5}$ (in units of the gravitino mass). 

This can have important implications for the standard two-step moduli stabilisation procedure where in the first step the axion-dilaton and the complex structure moduli are fixed at tree-level, and in the second step the  K\"ahler moduli are stabilised by subleading corrections while keeping the axion-dilaton and the complex structure moduli fixed at their tree-level VEV. The K\"ahler moduli stabilised by non-perturbative effects, as the volume mode in KKLT \cite{Kachru:2003aw} or as blow-up modes in LVS \cite{Balasubramanian:2005zx}, typically have masses slightly above the gravitino mass, while the K\"ahler moduli fixed by perturbative corrections, as bulk moduli in LVS \cite{Cicoli:2008va,Cicoli:2016chb} or in purely perturbative stabilisation schemes \cite{Berg:2005yu,Antoniadis:2019rkh}, are lighter than the gravitino. Given that in many cases we found numerically that some complex structure moduli are lighter than the gravitino mass, these fields would definitely be lighter than all K\"ahler moduli fixed by non-perturbative effects, and potentially lighter than those stabilised perturbatively. This result can therefore in principle invalidate a two-step procedure to stabilise the moduli where the complex structure moduli are integrated out before addressing K\"ahler moduli stabilisation. However this concern can be alleviated by noting that the tree-level K\"ahler potential (\ref{eq:TreeLevKP}) factorises, and so the two sectors do not mix at tree-level. Of course, the validity of the two-step moduli stabilisation procedure would have to be investigated in detail in any specific model.

Another important phenomenological aspect to consider is the cosmological moduli problem (CMP). The successes of Big-Bang Nucleosynthesis require to have moduli masses above about $30$ TeV \cite{Banks:1993en, deCarlos:1993wie}. Thus,  moduli masses well below the gravitino masses require a very large value of the gravitino mass. This, in turn, would typically imply a very large value of the soft masses for the supersymmetric partners in the visible sector, unless sequestering is at play \cite{Blumenhagen:2009gk,Aparicio:2014wxa}. In this context, it is important to keep in mind that the CMP bound assumes that the moduli suffer an initial displacement
of $\mathcal{O}(M_{\rm pl})$ which has to be checked for explicit models. In particular, as pointed out in \cite{Conlon:2007gk}, this is not expected to be the case for the axio-dilaton and the complex structure moduli which should therefore cause no CMP even if they are lighter than the gravitino. In fact, the potential for the complex structure moduli is steeper than the one for the K\"ahler moduli, and so the former are expected to be trapped very close to their minimum in the early universe without experiencing large displacements. Note moreover that large initial displacements in the complex structure moduli directions would destabilise the K\"ahler moduli due to the $\mathcal{V}^{-2}$ prefactor of the flux-generated scalar potential. Again, detailed investigations in specific models would be important.

\begin{figure}[t!]
\centering
  \includegraphics[width=1\linewidth]{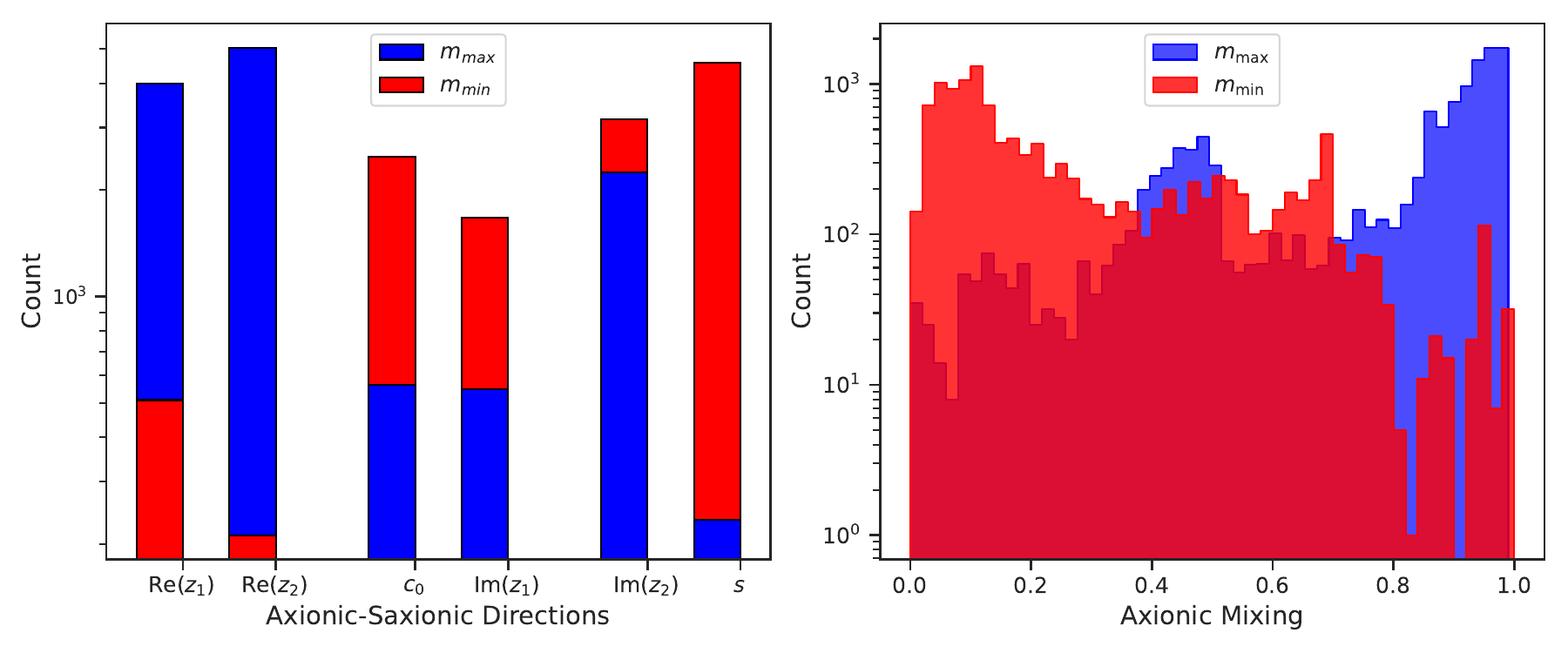}
\caption{Mass mixing between axions and saxions for dataset B. \emph{Left:} For each modulus we show the number of flux vacua where the mass eigenstate corresponding to the maximal and minimal mass eigenvalues is mainly aligned along that modulus. \emph{Right:} Number of flux vacua where the mass eigenstate corresponding to the maximal and minimal mass eigenvalues has a given percent mixing with the axionic directions.
}
\label{fig:masses}
\end{figure}

Fig. \ref{fig:masses} exhibits more detailed properties on the masses, in particular the mixing between axions and saxions while 
going to the mass eigenstates for dataset B. Our analysis focuses on studying the alignment of mass eigenstates with the \emph{axionic} directions corresponding to the real parts of the moduli\footnote{Note that this
nomenclature is somehow non-standard since we use the word \emph{axions} to denote fields which do not appear in the perturbative K\"ahler potential even if their shift symmetry is broken by fluxes in $W$.}. Interestingly, Fig. \ref{fig:masses} implies that the axionic directions tend to be steeper then the saxionic ones, and that the dilaton $s$ tends to be the lightest mode. This might have important implications for phenomenology. We plan to dedicate future work to explore the genericity of these features in the Type IIB flux landscape.

\section{Conclusions}
\label{sec:conclusions}

In this work we developed an algorithm and used targeted numerical methods to perform deep explorations of flux vacua in Type IIB flux compactifications. The constraints and algorithms developed in this work represent a significant advancement in the systematic study of flux vacua. We derived novel bounds on flux vectors $f$ and $h$, enabling efficient and targeted construction of vacua within specific regions of moduli space. These constraints are rooted in the eigenvalues of the ISD matrix and account for the intricate structure of the flux landscape, refining earlier approaches by introducing stricter bounds that reduce irrelevant flux sampling. 

Our algorithm leverages these constraints to efficiently generate consistent flux configurations, employing a systematic approach rather than relying on random sampling. By incorporating methods such as rounding continuous fluxes to integers and employing modular symmetries to remove redundant solutions, the algorithm ensures both thoroughness and computational efficiency. Compared to earlier methods, such as those based on random flux generation or restrictive parametrisations, our approach captures a far greater fraction of viable vacua. Its universality also makes it applicable to models beyond the two-moduli setup explored in the bulk of the paper, providing a robust framework for analysing diverse compactifications while retaining computational feasibility.

Focusing on a two-moduli model at large complex structure, we studied specific regions of moduli space, making use of the \texttt{JAXVacua} framework \cite{Dubey:2023dvu} to construct flux configurations, solve $F$-flatness conditions, and investigate phenomenological properties. We found local deviations in the density for the number of vacua and also deviations with the scale of the number of vacua with $N_{\text{max}}$ -- these findings highlight the limitations of statistical approaches. These discrepancies stem from moduli space geometry, flux constraints, and ISD sampling. Comparing numerical and statistical results validates our methods and underscores the need for refined analytic predictions in specific regions.

Further, we identified intriguing patterns in the distributions of the flux superpotential $W_0$ and moduli masses. The distribution of $W_0$ in the complex plane exhibited symmetry-breaking features, circular arcs, and voids, attributed to flux hierarchies and gauge-fixing effects. Our vacua included examples with low $|W_0|$, consistent with statistical prediction~\cite{Denef:2004ze}, further validating the framework. Furthermore, the distribution of $W_0$ in the complex plane sheds light on the global structure of the landscape, revealing patterns that may guide future model-building efforts. Additionally, we characterised mass hierarchies, revealing significant ranges of mass scales and notable mixing between axionic and non-axionic directions, with implications for moduli stabilisation, supersymmetry breaking, and de Sitter uplift scenarios. For instance, the relative scale of the gravitino mass compared to the moduli masses impacts stabilisation mechanisms and the viability of de Sitter uplift scenarios. These results underscore the utility of an exhaustive numerical approach in bridging the gap between theoretical predictions and observable phenomenological quantities.

Our findings highlight key directions for future research. Extending these methods to non-supersymmetric vacua is a promising avenue, although the absence of the ISD condition poses a significant challenge. Developing techniques to explore critical points of the scalar potential could yield insights into broader flux configurations, including potential de Sitter vacua from $F$-term uplifts.

The observed hierarchies among moduli masses and the intricate patterns in the superpotential distributions warrant a deeper investigation. What mathematical structures or symmetries underlie these distributions? Are they generic to certain classes of compactifications, or do they emerge from specific flux configurations? Understanding the origin of these features could shed light on the interplay between moduli stabilisation, phenomenological parameters, and the structure of the landscape. This analysis could also clarify how these hierarchies influence physical properties, such as the supersymmetry-breaking scale and the viability of de Sitter vacua.

These investigations aim to deepen our understanding of the interplay between geometry, fluxes, and vacuum structure in the string landscape. They also bridge the gap between statistical predictions and explicit constructions, offering a detailed view of local properties and global trends. This work provides a robust framework for targeted exploration of the string landscape and advances our ability to relate its rich mathematical structure to observable physics. Future research can refine theoretical models and pursue applications in high-energy physics and cosmology. By integrating data-driven methods with phenomenological constraints, we aim to inspire new directions in model building, advancing both theoretical understanding and practical applications.

\section*{Acknowledgements}

We would like to thank Thomas Grimm, Arthur Hebecker, Liam McAllister and Erik Plauschinn for interesting discussions. AC would like to thank Keshav. 
SK's work has been partially supported by STFC
consolidated grants ST/T000694/1 and ST/X000664/1. This article is based upon work from COST Action COSMIC WISPers CA21106, supported by COST (European Cooperation in Science and Technology).
This work made use of the open source software 
\texttt{CYTools}~\cite{Demirtas:2022hqf},
\texttt{jax}~\cite{jax2018github},  
\texttt{matplotlib}~\cite{Hunter:2007},  
\texttt{numpy}~\cite{harris2020array1},
\texttt{scikit-learn}~\cite{scikit-learn},
\texttt{scipy}~\cite{2020SciPy-NMeth1},
and \texttt{seaborn}~\cite{Waskom2021}. 

\appendix

\section{Details on integral computations}
\label{sec:appendix}

In this appendix we cover important details of numerically/analytically computing various integrals appearing in this work for counting vacua in a given region of moduli space as given by Eq.~\eqref{eq:ntot} in the main text and also the count on vacua with low $|W_0|$ given by Eq.~\eqref{eq:nsup}.

\subsection{Ingredients for the integrals}
\label{sec:integrations}

We start by establishing our conventions for the topological data used in the expression \eqref{eq:ntot} for $\cN_{\text{stat}}$. We define the rescaled Yukawa couplings in orthonormal frame as
\begin{equation}
    F_{IJK}=-\I (e^{\,\,a}_{I})(e^{\,\,b}_{J})(e^{\,\,c}_{K})y_{abc}\mathrm{e}^{{K}_{\text{cs}}}\, ,
\end{equation}
here, $I, J, K,\dots $ are the orthonormal frame indices and $a, b, c, \dots$ are special coordinate indices. The Yukawa couplings $y_{abc}\equiv\partial^{3}_{abc}({F_{\text{pert}}+F_{\text{inst}}})$ receive instanton corrections that are suppressed in the LCS limit, resulting in 
\begin{equation}
    y_{abc} = \widetilde{\kappa}_{abc} + \cO(\varepsilon)\, ,
\end{equation}
where $\widetilde{\kappa}_{abc}$ are the triple intersection numbers coming from $\partial^{3}_{abc}F_{\text{pert}}$. 
In evaluating various integrals, $e^{\,\,\, a}_{I}$ is used to transform quantities from the special coordinates to the orthonormal frame and is defined in terms of the complex structure moduli space metric $g^{a\Bar{b}}_{\text{cs}}$ as 
\begin{equation*}
    g^{a\Bar{b}}_{\text{cs}}= e^{\,\,a}_{I}\delta^{I\Bar{J}} e^{\,\,\,\Bar{b}}_{\Bar{J}}\, .
\end{equation*}     
Having described our conventions and topological data of the underlying compactification manifold, we now turn to computing these integrals.

\subsection{Number of vacua: statistics}
\label{app:number}

We present the details for evaluating the integral in \eqref{eq:ntot} for the case of two complex structure moduli. Let us define
\begin{align}
\label{eq:nstatAppendix}
I_{\text{tot}} \equiv \pi^{-6}\hspace{-.5cm}\int\limits_{\mathcal{M_{\tau}}\times\mathcal{M}_{\text{cs}}}\hspace{-0.5cm}\,\mathrm{d}\tau \,\mathrm{d}\bar{\tau} \prod_{i=1}^{h_{-}^{(1,2)}=2}\!\!\!\,\mathrm{d}z^{i}\,\mathrm{d}\bar{z}^i \, \text{det}g \int d^2Xd^{4}Z \exp(-|X|^2-|Z|^2)|X|^2 \bigg|\det
\begin{pmatrix}
A & B \\
C & D 
\end{pmatrix}\bigg|
\end{align} 
with\footnote{The bar over variables indicates the complex conjugate.}
\begin{align*} 
A = \bar{D} & = \delta^{IJ}\Bar{X}-\frac{\Bar{Z}^I Z^J}{X}\, , \quad B = \Bar{C} = {F}_{IJK}\Bar{Z}^K\,.
\end{align*}
There are 6 complex (12 real) integration variables: $X,\, Z^1,\, Z^2,\, \tau,\, z^1,\,$and $z^2$. The integral over $\tau= c_0 + \I s$, denoted by $I_{\tau}$, is factored from the total vacua integral in equation~\eqref{eq:nstatAppendix} since the vacua density $\rho(z)$ in \eqref{eq:den} does not depend on $\tau$.
For vacua with $\tau $ inside the complete fundamental domain, $I_{\tau }$ is given by the standard moduli space integral\footnote{Considering the metric $g_{\tau\Bar{\tau}} = \frac{1}{(\tau-\Bar{\tau})^2}$, we get $ \text{vol}(\mathcal{M}_{\tau})  =  \int_{-1/2}^{1/2} \mathrm{d}c_0 \int_{\sqrt{1-c_0^2}}^{\infty} \mathrm{d}s \frac{1}{(2s)^2} = \frac{\pi}{12} $} $\text{vol}(\mathcal{M}_{\tau}) \equiv \int_{\mathcal{M}_{\tau}} \mathrm{d}^2 \tau \, g_{\tau\Bar{\tau}} = \pi/12$.
To simplify the remaining integral, we first observe that the integrand in \eqref{eq:nstatAppendix} is invariant under the following phase transformations:
$X\rightarrow X\mathrm{e}^{i\theta}$, $Z^I\rightarrow \mathrm{e}^{i\beta}Z^I$.
The invariance of the determinant of the block matrix 
\begin{equation}
 \det\begin{pmatrix}
A & B \\
C & D 
\end{pmatrix} = \det(D)\det(A-BD^{-1}C)
\end{equation}
can be understood by examining the transformations of $A$, $B$, $C$, and $D$. 
Only the relative phase between $Z_1$ and $Z_2$, denoted by $\alpha$, is relevant under such transformations. After eliminating $\tau$ and the absolute phases \footnote{We go to polar coordinates for $X, Z^1$ and $Z^2$ and integrate out the absolute phase $\theta$ and $\beta$, giving $2\pi$ factor each.} of $X$ and $Z^1$ and $Z^2$. We are left with 8 real variables $|X|, |Z^1|, |Z^2|, \alpha, \text{Re}(z^i), \text{Im}(z^i)$ for $i=1,2$.
Since we consider the \emph{large complex structure} limit of the moduli space where the instanton effects are negligible, the existence of approximate shift symmetries $\text{Sp}(6, \mathbb{Z})$ allows us to restrict real parts of the complex structure moduli to $[-0.5, 0.5]$. Hence, the Re${(z^i)}$ can be integrated\footnote{Moreover, ${F}_{IJK}\Bar{Z}^{K}$ does not affect Re$(z^i)$ in the \emph{LCS} limit.}  trivially giving unity for each complex structure modulus. The remaining 6 real variables are $|X|, |Z^1|, |Z^2|, \alpha, \text{Im}(z^1),$ and $\text{Im}(z^2)$.
Employing the above simplifications, the 6-dimensional integral becomes:
\begin{multline}
\label{eq:nmonteAppendix}
\frac{(2\pi)^2}{\pi^6} \cdot I_{\tau} 
\int_{\mathcal{M}_{LCS}} (\text{det}g)_{\text{cs}} \, \,\mathrm{d}\text{Im}(z^1) \, \,\mathrm{d}\text{Im}(z^2) 
\int \,\mathrm{d}|X| \, \mathrm{d}|Z_1| \, |Z_1|  \,\mathrm{d}|Z_2| \, |Z_2| \\
\int_0^{2\pi} \,\mathrm{d}\alpha \, 
\exp(-|X|^2 - |Z_1|^2 - |Z_2|^2) |X|^3 \\
\biggl[ 
| \text{det}(|X|, |Z_1|, |Z_2|, \alpha)| 
+ | \text{det}(|X|, |Z_1|, |Z_2|, -\alpha)| 
\biggr] (2\pi - \alpha).
\end{multline}
The ranges for dummy variables $|X|, |Z_1|$, and $|Z_2|$ in \eqref{eq:nmonteAppendix} are chosen uniformly in $[0,5]^3$, as the term $|X|^3\exp(-|X|^2 - |Z_1|^2 - |Z_2|^2)$ is rapidly decaying and effectively supported in this range.

The integral in \eqref{eq:nmonteAppendix} is evaluated using Monte-Carlo methods. The results for the regions in the four datasets described in Tab.~\ref{tab:summary} and the corresponding values of the statistical predictions for the total number of vacua are presented in Tab.~\ref{tab:appTab}.

\begin{table}[t!]
    \centering
    \begin{tabular}{|c||c|c|}
         \hline
         & &  \\[-1.3em]
         Dataset & $I_{\text{tot}}$ & $\cN_{\text{stat}}$ \\[0.2em]
         \hline
         \hline 
         & &  \\[-1.2em]
         A & $5.6607\times 10^{-5} \pm 2.1964 \times 10^{-8} $ & 7,472,987 $\pm$ 2,899 \\[0.3em]
         \hline
         & &  \\[-1.2em]
         B & $1.7966\times 10^{-4}\pm 8.9365 \times 10^{-8}$ & 15,353 $\pm$ 7\\[0.3em]
         \hline
         & &  \\[-1.2em]
         C & $2.0430 \times 10^{-3} \pm 1.0117 \times 10^{-6}$& 269,427,663 $\pm$ 1,33,570\\[0.3em]
         \hline
         & &  \\[-1.2em]
         D & $3.3778\times 10^{-4} \pm 1.5431 \times 10^{-7}$& 451,038,133 $\pm$ 2,06,057\\[0.3em]
         \hline 
    \end{tabular}
\caption{Values of the integral $I_{\text{tot}}$ defined in \eqref{eq:nstatAppendix} and the corresponding total number of vacua $\cN_{\text{stat}}$ for the four datasets of Tab. \ref{tab:summary}. The errors in the values are at most $0.1$ percent of the average value.}
\label{tab:appTab}
\end{table}

\subsection{Vacua with small superpotential}
\label{sec:smallWstat}

We now briefly present the details to evaluate the integral in  \eqref{eq:nsup} which determines the number of vacua with $|W_0|\ll 1$. Let us
define:
\begin{align}
\label{eq:lowsupAppendix}
          I_{|W_0|} \equiv  \int_{\mathcal{M}} \,\mathrm{d}^6z \,\text{det}(g) \,\mathrm{e}^{2{K}_{\tau}+2{K}_{\text{cs}}}{F}_{abc}\Bar{{F}}^{abc}.
\end{align}
As before, this integral factorises into a $\tau$ piece and complex-structure piece. For $\tau = c_0 + \I s$,  the axio-dilaton integral can be performed explicitly inside a subregion of the fundamental domain of $\tau$  and reads: 
\begin{equation}
    I_{|W_0|,\tau} = \int^{+1/2}_{-1/2} \,\mathrm{d}c_0 \int^{s_2}_{\sqrt{1-c^2_0}} \,\mathrm{d}s \frac{1}{(2s)^4}. 
\end{equation}
The complex-structure contribution in \eqref{eq:lowsupAppendix} after trivially integrating\footnote{Each gives a factor of $1.$} the Re$({z^i}) \in [-0.5, 0.5]$ simplifies, for e.g. dataset A, to
\begin{equation}
    I_{|W_0|, \text{cs}} = 
    \int_{2}^{3} \mathrm{d}\text{Im}(z^1) \int_{2}^{3} \mathrm{d}\text{Im}(z^2)\,\text{det}(g_{\text{cs}}) \, \mathrm{e}^{2{K}_{\text{cs}}}{F}_{abc}\Bar{{F}}^{abc},
\end{equation}
where the integral $I_{cs}$, depending only on Im${(z^i)}$ and Calabi-Yau data, can be evaluated numerically using Monte-Carlo methods. The values of $I_{|W_0|}$ for the dataset described in Tab. \ref{tab:summary} are collected in Tab.~\ref{tab:appTab2}.

\begin{table}[t!]
    \centering
      \begin{tabular}{|c||c|c|}
         \hline
         &    \\[-1.3em]
         Dataset  & $I_{|W_0|}$ \\[0.2em]
         \hline
         \hline 
         &     \\[-1.2em]
         A  & $8.8125 \times 10^{-5} \pm 3.1787 \times 10^{-9}$\\[0.3em]
         \hline
         &   \\[-1.2em]
         B & $2.9692 \times 10^{-4} \pm 2.0801 \times 10^{-8}$\\[0.3em]
         \hline
         &    \\[-1.2em]
         C & $2.8246 \times 10^{-3}\pm 1.3107 \times 10^{-6}$\\[0.3em]
         \hline
         &   \\[-1.2em]
         D & $5.5812 \times 10^{-4} \pm 9.8383 \times 10^{-8}$\\[0.3em]
         \hline 
    \end{tabular}
\caption{Values of the integral $I_{|W_0|}$  defined in \eqref{eq:lowsupAppendix} for the datasets summarised in Tab.~\ref{tab:summary}. }
\label{tab:appTab2}
\end{table}

\bibliographystyle{utphys}
\bibliography{Literatur}

\end{document}